# Environmental and Safety Impacts of Vehicle-to-Everything Enabled Applications: A Review of State-of-the-Art Studies

Jianhe Du, Kyoungho Ahn, Mohamed Farag, Hesham A. Rakha, *Fellow, IEEE*

*Abstract*—With the rapid development of communication technology, connected vehicles (CV) have the potential, through the sharing of data, to enhance vehicle safety and reduce vehicle energy consumption and emissions. Numerous research efforts have been conducted to quantify the impacts of CV applications, assuming instant and accurate communication among vehicles, devices, pedestrians, infrastructure, the network, the cloud, and the grid, collectively known as V2X (vehicle-to-everything). The use of cellular vehicle-to-everything (C-V2X), to share data is emerging as an efficient means to achieve this objective. C-V2X releases 14 and 15 utilize the 4G LTE technology and release 16 utilizes the new 5G new radio (NR) technology. C-V2X can function without network infrastructure coverage and has a better communication range, improved latency, and greater data rates compared to older technologies. Such highly efficient interchange of information among all participating parts in a CV environment will not only provide timely data to enhance the capacity of the transportation system but can also be used to develop applications that enhance vehicle safety and minimize negative environmental impacts. However, before the full benefits of CV can be achieved, there is a need to thoroughly investigate the effectiveness, strengths, and weaknesses of different CV applications, the communication protocols, the varied results with different CV market penetration rates (MPRs), the interaction of CVs and human driven vehicles, the integration of multiple applications, and the errors and latencies associated with data communication. This paper reviews existing literature on the environmental, mobility and safety impacts of CV applications, identifies the gaps in our current research of CVs and recommends future research directions. The results of this paper will help shape the future research direction for CV applications to realize their full potential benefits.

## I. INTRODUCTION

The transportation system is growing rapidly, with the current number of registered vehicles at 276 million, a total of 4.17 million miles of highways, and 3.23 trillion vehicle miles of travel (VMT)[1]. This gigantic system operates with a significant impact on the environment and traffic safety due to the huge number of vehicles and associated delays in traffic. According to *Forbes*, Americans lost nearly 99 hours, or an average of $1,377 per person in 2019 due to traffic crashes[2]. In addition, at least 38,800 people were killed in motor vehicle collisions[3]. The majority of these crashes were caused by human errors, such as distraction, driver inexperience, drowsiness, or speeding, all of which may be mostly avoided if warnings could be provided to drivers ahead of time. A report released by INRIX found that the economic impact of traffic congestion is both broad and complicated. Congestion cost New York $11 billion dollars, Los Angeles $8.2 billion dollars, and Chicago $7.6 billion dollars in 2019[4]. Safety and efficiency are not the only concerns regarding the traffic system; environmental impact has also become a major issue following the realization that climate change is caused by human activity. The U.S. Environmental Protection Agency (EPA) published the *Inventory of U.S. Greenhouse Gas Emissions and Sinks* in April 2021. This annual report summarizes the latest information on U.S. greenhouse gas emission trends from 1990 through 2019 [1]. The report identified the primary sources of greenhouse gas emissions in the U.S. and demonstrated that the transportation sector generated the largest share of greenhouse gas emissions, at 29%. The report also mentions that petroleum-based transportation fuel, including gasoline and diesel, is responsible

This work was jointly supported by Qualcomm and the University Mobility and Equity Center (UMEC). *Corresponding author: H.A. Rakha, hrakha@vt.edu.*

Jianhe Du is with the Virginia Tech Transportation Institute; 3500 Transportation Research Plaza, Blacksburg, VA 24061 USA (e-mail: jdu@vtti.vt.edu)

Kyoungho Ah is with the Virginia Tech Transportation Institute; 3500 Transportation Research Plaza, Blacksburg, VA 24061 USA (e-mail: kahn@vt.edu)

Mohamed Farag is with the Virginia Tech Transportation Institute; 3500 Transportation Research Plaza, Blacksburg, VA 24061 USA (e-mail: MFarag@vtti.vt.edu) and College of Computing and Information Technology, Arab Academy for Science, Technology, and Maritime Transport, Alexandria, Egypt (e-mail: mmagdy@aast.edu)

Hesham A. Rakha is with the Virginia Tech Transportation Institute; 3500 Transportation Research Plaza, Blacksburg, VA 24061 USA (e-mail: hrakha@vt.edu)

[1] Data source: https://www.statista.com. Due to the impact of COVID-19, all the numbers refer to 2019.
[2] https://www.forbes.com/sites/greggardner/2020/03/09/americans-lost-nearly-100-hours-to-traffic-in-2019-says-inrix/?sh=541e6b57d5c8
[3] https://safer-america.com/car-accident-statistics/
[4] https://inrix.com/scorecard/

for over 90% of the total transportation sector's fuel consumption. Total transportation emissions increased from 1990 to 2019 due to increased travel demand. In particular, the total vehicle miles traveled (VMT) in the U.S. increased by 48% during that period of time due to population growth, economic growth, urban sprawl, and periods of low fuel prices [2]. The EPA recommends reducing transportation-related emissions by switching to alternative fuel sources; improving fuel efficiency with advanced design, materials, and technologies; improving operating practices; and reducing travel demands.

Connected vehicle (CV) technology enables equipped vehicles to connect with other vehicles, roadway infrastructure, pedestrians, bicycle riders, and other devices through advanced wireless communication to improve roadway safety, travel efficiency, energy efficiency, and reduce vehicle emissions. Not only can CV applications increase throughput and mobility but may also, by preventing human errors, reduce vehicle crashes. We expect that CV technology will significantly increase the mobility and safety of the transportation system and reduce greenhouse gas emissions by using advanced technologies and improving transportation operational practices.

Vehicle-to-everything (V2X) technology represents communication between a vehicle and any object that can communicate with the CV. The seven types of vehicle connectivity [3] include vehicle-to-infrastructure (V2I), vehicle-to-vehicle (V2V), vehicle-to-network (V2N), vehicle-to-cloud (V2C), vehicle-to-pedestrian (V2P), vehicle-to-device (V2D), and vehicle-to-grid (V2G). Collectively these are known as V2X. V2I is a communication protocol that bi-directionally communicates information between the vehicle and the road infrastructure. V2V enables equipped vehicles to exchange real-time data with other equipped vehicles. V2N allows vehicles to communicate with the network, including the V2X management system. V2C enables equipped vehicles to offer bi-directional data exchange with the cloud, including digital assistants and the Internet of Things. V2P includes communication with road users, including pedestrians, people using wheelchairs, people riding bicycles, and people using other mobility devices. V2D allows equipped vehicles to exchange information with any smart device such as smartphones, tablets, and other wearable devices. V2G is a communication with the smart electric grid in support of electrification of the transportation system for equipped vehicles.

This paper summarizes the potential mobility, environmental, and safety benefits of V2X-enabled applications, with a focus on cellular V2X (C-V2X) applications, based on recently published literature. The potential benefits can vary based on the application type, study design, vehicle type, test location or network type, utilized energy/emission model, and market penetration rates (MPRs) of CVs. The conclusions drawn by these previous studies confirmed that, through an efficient exchange of information among vehicles, infrastructures, network, devices, and other participating parts of the transportation system, the transportation system can operate more efficiently and with reduced emissions. CVs and C-V2X enabled applications, by providing driving directions, speed advice, and acceleration or deceleration suggestions, can help in reducing delays, increasing vehicle throughput, reducing greenhouse gas emissions, and improving fuel and energy efficiency, while at the same time significantly decreasing the number of vehicle crashes. Such benefits can be significant when the MPR of CVs reaches a certain level. Furthermore, not only do CVs benefit from V2X communications, but other non-connected vehicles also benefit from the CVs' improved efficiency, safety, and decreased emissions, because when CVs use the transportation system more efficiently and safely, the system has more redundancy for non-connected vehicles.

This paper categorized three CV studies into three areas: 1) applications ignoring communication system constraints; 2) applications modeling the mutual dependencies of the transportation and communication systems; and 3) applications considering communication system constraints. The first area includes research that ignores the communication system constraints. In this category, researchers concentrated on modeling the benefits and impacts of CV applications assuming that the communication among vehicles and infrastructures is perfect without any delays or errors. In addition, the drivers of the CVs are assumed to follow the guidance or suggestions provided by the CV applications completely. The second category includes efforts that attempt to integrate transportation and communication system modeling for the evaluation of CV applications accounting for the communication system constraints. The third includes research that evaluated applications considering the communication system constraints. Due to the limited communication bandwidth, there will be congestion in transmitting signals and data in the CV system, especially during peak demand periods when there are overwhelming data that need to be transmitted. This section focuses on previous studies that accounted for communication system constraints when evaluating the impacts of CV applications.

## II. APPLICATION EVALUATIONS IGNORING COMMUNICATION SYSTEM CONSTRAINTS

The majority of existing literature on CV applications does not consider communication system constraints on system performance. Specifically, these studies assume that the data related to vehicle locations, vehicle kinematics, infrastructure status, traffic controls, as well as travelers of all the modes in the transportation network, can be transmitted to each agent in the system instantaneously with no loss or delay in data packets. The applications that build on V2X communication can be sorted into four categories: system/network level applications, freeway applications, signal-free intersection applications, and signalized corridor applications, as described in the following sections.

*A. Network Applications*

This section summarizes the various efforts reported in the literature quantifying the network-wide impacts of CV applications. As summarized in Table I, we found a limited number of studies focusing on network-wide impacts of CV applications. All studies entailed evaluating these applications in a virtual traffic simulation environment given that actual field implementations are cost prohibitive.

A research study by Ahn et al. developed and evaluated an Eco-Cooperative Automated Control System (Eco-CAC) system that integrates vehicle control strategies with connected automated vehicle (CAV) applications. The developed system was tested on a large-scale network in downtown Los Angeles, CA, considering a combination of internal combustion engine vehicles (ICEVs), hybrid electric vehicles (HEVs), and battery electric vehicles (BEVs) in a microscopic traffic simulation environment for three different demand levels: no congestion, mild congestion, and heavy congestion. The Eco-CAC system includes an Eco-router, a speed harmonization (SPD-HARM) controller, an Eco-CACC-I controller (also known as green light optimized speed advisory [GLOSA]), and an Eco-CACC-U controller (also known as vehicle string control or platooning). The study found that the Eco-CAC system reduces fuel consumption by up to 16.8%, energy consumption by up to 36.9%, travel time by up to 21.7%, total delay by up to 43.1%, stopped delay by up to 68.7%, and $CO_2$ emissions by up to 16.8%. The study also found that different vehicle types generate different benefits. For example, the Eco-CAC system reduces fuel consumption, travel time, total delay, stopped delay, and $CO_2$ emissions for ICEVs in heavily congested conditions. The benefits of the controllers mostly increase as the CAV MPR increases. However, the controllers negatively affect ICEV fuel consumption, travel time, total delay, stopped delay, and $CO_2$ emissions for no or mild congestion. For BEVs, the study found that the Eco-CAC system improves the energy consumption but negatively affects travel time, total delay, and stopped delay for all congestion levels. The study also tested the Eco-CAC system considering both the current and a future vehicle composition on the LA network and found that the Eco-CAC system effectively reduces fuel and energy consumption, travel time, total delay, and stopped delay in heavily congested conditions for both vehicle compositions. However, the study found that different vehicle compositions produced different results. Specifically, the maximum energy consumption savings of BEVs (36.9% saving) for the current vehicle composition were observed at a 10% CAV MPR in mild congestion, while the maximum savings for the future vehicle composition (35.5%) were observed at a 50% CAV MPR with no congestion. The study found that the developed Eco-CAC system effectively reduces the fuel and energy consumption, travel time, total delay, stopped delay, and $CO_2$ emissions for ICEVs, BEVs, and HEVs for specific scenarios. Additionally, the effectiveness of the Eco-CAC system and the controllers depends on the traffic conditions, including the level of traffic congestion, the network configuration, the CAV MPR, and the vehicle composition [4].

The Netherlands Organization for Applied Science (TNO) investigated various CV use cases. The study reviewed available literature on the environmental impact of CV use cases and the study performed simulation studies using a microscopic emission calculation tool (EnViVer) to quantify the potential environmental impacts of CV applications. EnViVer was developed using the VISSIM traffic microsimulation software and VERSIT+, a vehicle emission simulation tool. The identified mechanisms that were taken into consideration include reduction of trips; reduction and departure time shift; mode shift, reduction of vehicle dynamics, and the powertrain operation. The results showed that an eco-driving technique that reduces vehicle stops is very beneficial, generating a $CO_2$ reduction in the range of 13–45%. An eco-driving technique that reduces deceleration and acceleration shows benefits as well (3–7% improvement). The study found that cooperative adaptive cruise control (CACC) reduces vehicle emissions by 6% compared to ACC [5].

Olia et al. studied the potential impacts of CV applications using PARAMICS. The simulation testbed was a network in the north of Toronto. The simulation results showed that CVs were able to navigate through multiple routing options and therefore could save up to 58% in travel time at an MPR of 50% on certain corridors. As for improving safety indicators, increasing the MPR of CVs can improve the safety index (the probability of incidents) by up to 45%. The Comprehensive Modal Emission Model (CMEM) was used to estimate emission factors. A reduction of 30% in $CO_2$ emissions was identified [6].

In a study by Vahidi et al., the authors reviewed the energy saving potential of CAVs based on first principles of motion and optimal control theory. The study found that connectivity to other vehicles and infrastructure allows better anticipation of upcoming events, such as hills, curves, slow traffic, state of traffic signals, and movement of neighboring vehicles. The review paper concluded that automation allows vehicles to adjust their motion more precisely in anticipation of upcoming events to save energy, and that cooperative driving can further increase a group of vehicles' energy efficiency by allowing them to move in a coordinated manner. The study concluded that CAVs' energy efficient movement could have a harmonizing effect on mixed traffic, leading to additional energy savings for neighboring vehicles [7].

A study by Rahman et al. investigated the safety impact of CVs and lower-level automation (CVLLA) using simulation. Two features were tested as CVLLA: automated braking and lane keeping assistance. Segment and intersection crash risks were estimated using surrogate safety assessment modeling techniques. The results showed that both CV and CVLLA technology significantly reduced conflict frequency. Higher MPRs of CVs generate higher benefits, where the maximum improvement was found to be at 100% MPR. However, at least 40% MPR is needed to achieve the safety benefits of reduced intersection crash risks and 30% is needed for reduced segment crash risks [8].

TABLE I
V2X-ENABLED NETWORK APPLICATIONS

| Year, Title | Leading Author, Institute | CV Application | Road type | Comm. type | Topic area | Estimated Benefits | Simulation/ Field Test | Modeling Environment |
|---|---|---|---|---|---|---|---|---|
| 2021, Evaluating an Eco-Cooperative Automated Control System (Eco-CAC) | Ahn, K – VTTI | Various – Eco-routing, SPD-HARM, Eco-approach and departure, Platooning | Network | N/A | Mobility, Energy, Environment | Reduced fuel consumption by up to 16.8%, BEV energy consumption by up to 36.9%, travel time by up to 21.7%, total delay by up to 43.1%, stopped delay by up to 68.7%, and $CO_2$ emissions by up to 16.8%. | Simulation | INTEGRATION, VT-CPFM, VT-CPEM |
| 2020, Environmental Benefits of C-V2X | TNO | Various – CACC, Eco-driving, Intelligent intersection | Arterial and Freeway | N/A | Environment | $CO_2$ reduction in the range of 13-45%. | Simulation | EnViVer (VISSIM and VERSIT+) |
| 2016, Assessing the Potential Impacts of Connected Vehicles: Mobility, Environmental, and Safety Perspectives | Olia, A – McMaster University | Routing and safety | Network | N/A | Mobility, Safety, Environment | Reduced travel time by 37%, reduced emissions by 30% and improved safety indicators by 45%. | Simulation | Paramics, CMEM |
| 2018, Energy saving potentials of connected and automated vehicles | Vahidi, A – Clemson | Review paper – Various | Network | N/A | Environment | This paper summaries the benefits of coordinated and smoother motion of CAVs, in terms of car following, lane changing, and intersection control on energy savings and environment impacts. The authors concluded that from previous literatures, the savings on energy can range from 3% to 20% on varied facilities. | N/A | N/A |
| 2019, Safety benefits of arterials' crash risk under connected and automated vehicles | Rahman, MS – University of Central Florida | Automated braking and lane keeping assistance | Arterial | N/A | Safety | Travel time saving between 59% to 84%. | Simulation | VISSM |
| 2021, Multi-objective Eco-Routing Model Development and Evaluation for Battery Electric Vehicles | Ahn, K – VTTI | Eco-Routing | Network | N/A | Energy | The multi-objective routing reduced BEV energy consumption up to 14.2% and ICEV fuel consumption up to 10.6%. | Simulation | INTEGRATION, VT-CPFM, VT-CPEM |
| 2020, Development of a Connected Vehicle Dynamic Freeway Variable Speed Controller | Abdelghaffar, H – VTTI | SPD-HARM | Network | N/A | Mobility, Environment | 12.17% reduction in travel time, 20.67% reduction in total delay, 2.6% fuel consumption saving, $CO_2$ emission savings of 3.3% | Simulation | INTEGRATION, VT-CPFM |

A study by Ahn et al. developed a multi-objective eco-routing algorithm (eco- and travel time-optimum routing) for BEVs and ICEVs in a CV environment and investigated the network-wide impacts of the multi-objective Nash optimum (user equilibrium) traffic assignment on a large-scale network. Unlike ICEVs, BEVs are more energy efficient on low-speed arterial trips compared to highway trips. The authors demonstrated that different energy consumption patterns require different eco-routing strategies for ICEVs and BEVs. This study found that single objective eco-routing could significantly reduce BEVs' energy consumption but also significantly increased their average travel time and thus the authors developed a multi-objective routing model (eco- and travel time-routing) to improve both energy and travel time measures. The simulation study found that multi-objective routing could reduce BEVs' energy consumption by 13.5%, 14.2%, 12.9%, and 10.7%, as well as ICEVs' fuel consumption by 0.1%, 4.3%, 3.4%, and 10.6% for "not congested, "slightly congested," "moderately congested," and "highly congested" conditions, respectively. The study also found that multi-objective user equilibrium routing reduced the average vehicle travel time by up to 10.1% compared to the standard user equilibrium traffic assignment for highly congested conditions, producing a solution closer to the system optimum traffic assignment. The study concluded that the multi-objective eco-routing strategy can reduce vehicle fuel/energy consumption effectively with minimum impacts on travel times for both BEVs and ICEVs [9].

In a study conducted by Abdelghaffar et al., a dynamic freeway speed controller based on sliding mode theory was developed and tested. The advantages of the sliding mode control are its simple design, global stability, and robustness that can address the discontinuity of the fundamental diagram due to the capacity drop. A SPD-HARM controller was developed in their study to dynamically identify bottlenecks and regulate the speeds of CVs. A decentralized phase split/cycle length controller was used to optimize all traffic signals in the network. The results showed that average travel time was reduced by 12%, total delay was reduced by 21%, and $CO_2$ emissions were reduced by 3.3%. In addition, the authors' results showed that freeways benefit from the controller more than arterials [10].

In summary, we found that the benefits of CV applications were highly dependent on the application type, network settings, MPRs, and test algorithms. There were a limited number of studies focusing on the benefits at a network/system level. More studies concentrated on either freeways only or signalized intersection corridors only. The following two sections will be used to illustrate these two categories individually.

*B. Freeway Applications*

Numerous studies investigated and quantified the impacts of CV-enabled applications on freeway sections, as summarized in Table II. Some of these studies concentrated on car-following and lane-changing behavior using a basic multi-lane freeway segment without any other complex roadway configurations. Some concentrated on studying the disturbance of merging or diverging traffic on a mainline freeway by analyzing freeway sections with on- and off-ramps. For example, in a study by Jang et al., the authors analyzed crash risks and estimated the safety benefits of the forward hazardous situation warning information presented by a Connected Intelligence Transport System (C-ITS) pre-deployment project for Korean freeways. C-ITS is composed of on-board units, roadside units (RSUs), and a traffic management center. The crash potential index (CPI) was adopted to quantify the crash benefits of CVs. A total of 700 CVs, including 400 buses, 264 trucks, and 36 sports utility vehicles, were instrumented with data acquisition systems and V2X communication devices. The study found that the average speed decreased by 10.2% and the time-to-collision (TTC) increased by 5.3% when warning information was provided in a CV environment. In addition, the achievable reduction in the CPI was approximately 20.7% due to the provision of warning information [11].

To study the effectiveness and network communication efficiency of connected (V2V and V2I) vehicular technologies in alerting motorists when they are approaching a hazardous zone, such as using CV technology to recommend a proper speed for travelers going through a low visibility area, Outay et al. compared the performance of V2V and V2I communications in an extensive computer simulation experiment. The authors adapted the iTetris platform for various scenarios—a baseline scenario, a deactivated alert scenario, a V2V cooperative alert scenario, and a V2I alert scenario. The authors also explored, via simulations, whether Cooperative Hazard Awareness and Avoidance systems, based on V2V and V2I communications can potentially contribute towards eco-driving by reducing $CO_2$ emissions. The results revealed that the alerting system based on V2I communication yields better message reception rates and better safety efficiency. The results also showed that the Cooperative Hazard Awareness and Avoidance system can reduce $CO_2$ emissions using SPD-HARM [12].

To explore the impacts of CACC on vehicle fuel efficiency in mixed traffic, Liu et al. conducted analyses at a freeway merge bottleneck. Their results showed that CACC string operation resulted in a maximum of 20% reduction in energy consumption compared to the human driver only case. At a 100% MPR, CACC-equipped vehicles consumed 50% less fuel than ACC vehicles without V2V communication and cooperation. At lower CACC MPRs, the vehicle fuel efficiency could be improved via use of a dedicated CACC lane or by implementing wireless connectivity on the manually driven vehicles. In the case of a CACC MPR of 40%, those strategies brought about a 15% to 19% capacity increase without decreasing vehicle fuel efficiency. The authors' results imply the importance of incorporating the V2V cooperation component into an automated speed control system and highlight the necessity of deploying CACC-specific operation strategies at lower CACC MPRs [13].

A study by Li et al. used simulation coding in MATLAB to test the effectiveness of reducing rear-end collisions in an I2V

system integrating variable speed limit (VSL) and ACC. The results showed that both the surrogate crash risk measures—time exposed time to collision (TET) and time integrated time to collision TIT—were reduced. VSL-only and ACC-only methods had a positive impact in reducing the TET and TIT values (reduced by 53.0 and 58.6% and 59.0 and 65.3%, respectively). The I2V system combined the advantages of both ACC and VSL to achieve the most safety benefits (reduced by 71.5 and 77.3%, respectively) [14].

A study by Rahman and Abdel-Aty attempted to evaluate longitudinal safety of CV platoons by comparing the implementation of managed-lane and all-lane CV platoons (for the same MPR) over a non-CV scenario. A high-level control algorithm of CVs in a managed-lane was proposed to form platoons with three joining strategies: rear join, front join, and cut-in join. Five surrogate safety measures—standard deviation of speed, TET, TIT, time exposed rear-end crash risk index, and sideswipe crash risk—were utilized as indicators for safety evaluation. The results showed that with CVs, the safety in the studied expressway significantly improved. Managed lane control produced better improvements compared to all-lane control for the same MPR [15].

Jin et al. used micro simulation to evaluate the mobility and environmental benefits of a CV application that involved a real-time optimal lane selection algorithm. On average, the travel time was reduced up to 3.8% and the fuel consumption was reduced by 2.2%. In addition, the reduction in emissions of criteria pollutants, such as CO, HC, $NO_x$ and PM2.5, ranged from 1% to 19%, depending on the different congestion levels of the roadway segment [16].

A study conducted by Guériau integrated multi-agent cooperative traffic modeling into the MovSim, a traffic simulator, to model complex interactions between cooperative vehicles and potentially among vehicles and infrastructure. The authors discussed some potentialities of C-ITS for traffic management with the methodological issues following the expansion of such systems. The operational goal of this work was to develop a decision-making tool validated in simulation conditions tailored to cooperative strategies. The model involved three layers that couple different dynamics to take into account information reliability while limiting traffic disturbances, and hence homogenize traffic flow. The results showed that at a 40% to 50% MPR, the connected environment had benefits in terms of homogenization and was able to avoid congestion for the merging traffic from on ramps and a scenario where a lane was closed [17].

Monteil et al. constructed a multi-agent framework that inherits knowledge from traffic theory. Their goal was to set up the modeling bricks of a cooperative auto-adaptive system with the assumption that modeling and technology uncertainties should affect the traffic physics and communications. A three-layer framework with physical, communication, and trust layers was used to achieve traffic flow homogenization. By modeling the trust as a function of distances, along with a communication layer and a physical layer, the authors conducted simulations using two lanes of freeway traffic with an entrance flow distribution from US-101 sample data. The results showed that the operation could decrease the speed variance and therefore the likeliness that traffic would fall into local congestion phenomena. The cooperation can also limit the impact of aggressive drivers (who oppose a global gain in acceleration when making a lane change). An increasing MPR reduces the impact of aggressive lane-changing behaviors, as it increases the stability domains and homogenizes traffic and reduces propagation of perturbations [18].

A study conducted by Liu et al. examined the impact of CACC string operations on vehicle speed and fuel economy on a 13-mile section of the SR-99 corridor near Sacramento, CA. The authors used simulation to evaluate the performance of the corridor under various CACC MPR scenarios and traffic demand levels. The CACC string operation was also analyzed when a vehicle awareness device (VAD) and CACC managed lane strategies were implemented. The results revealed that the average vehicle speeds increased by 70% when the CACC MPR increased from 0% to 100%. The highest average fuel economy, expressed in miles per gallon (MPG), was achieved under the 50% CACC scenario with an MPG at 27. This was 10% higher than the baseline scenario. However, when the CACC MPR was 50% or higher, the vehicle fuel efficiency only had minor increases. When the CACC MPR reached 100%, the corridor allowed 30% more traffic to enter the network without experiencing reduced average speeds. Results also indicated that the VAD strategy increased the speed by 8% when the CACC MPR was 20% or 40%, while there was a minor decrease in the fuel economy. The managed lane strategy decreased the corridor performance when implemented alone [19].

A study conducted by Ard et al. demonstrated the effectiveness of an anticipative car-following algorithm in reducing energy use of gasoline engine and electric CAVs. Without sacrificing safety and traffic flow, the authors implemented a vehicle-in-the-loop testing environment. Experimental CAVs were driven on a track, interacting with surrounding virtual traffic in real-time. The authors explored the energy savings in microsimulations. Model predictive control handled high level velocity planning and benefited from communicated intentions of a preceding CAV or estimated probable motion of a preceding human driven vehicle. A combination of classical feedback control and data-driven nonlinear feedforward control of pedals achieved acceleration tracking at the low level. The controllers were implemented in a robot operating system and energy was measured via calibrated OBD-II readings. The authors reported energy savings of 30% [20].

TABLE II
V2X-ENABLED FREEWAY APPLICATIONS

| Year, Title | Author, Leading institute | CV Application | Roadway Type | Comm. Type | Topic Area | Estimated Benefits | Simulation/ Field Test | Modeling Environment |
|---|---|---|---|---|---|---|---|---|
| 2020, Identification of safety benefits by inter-vehicle crash risk analysis using connected vehicle systems data on Korean freeways | Jang, J. – Hanyang University | Forward hazardous situation warning system | Freeway | Dedicated Short Range Communicat ions (DSRC) (1 Hz) | Safety | Reduction in the CPI was approximately 20.7 %; TTC increased by 5.3% | Field test | |
| 2019, V2v and V2I communications for traffic safety and $CO_2$ emission reduction: A performance evaluation | Outay, F. – Zayed University, UAE | Hazardous zone detection and alert system | Freeway | DSRC | Safety, Environment | TTC improved slightly; $CO_2$ improved by 5% | Simulation | iTetris platform (SUMO) |
| 2020, Freeway vehicle fuel efficiency improvement via cooperative adaptive cruise control | Liu, H. – UC Berkeley | CACC | Freeway | N/A | Mobility, Environment | A maximum of 20% reduction in energy consumption; Up to 49% of capacity increase | Simulation | NGSIM (VT-CPFM+MOVE S) |
| 2016, Reducing the risk of rear-end collisions with infrastructure-to-vehicle (I2V) integration of variable speed limit control and adaptive cruise control system | Li, Y. – Southeast University | Variable Speed Limit (VSL) and Adaptive Cruise Control (ACC) | Freeway | N/A | Safety | TET and TIT were reduced by 53% and 58.6% for VSL; 59% and 65.3% for ACC; 71.5% and 77.3% for combined VSL and ACC | Simulation | MATLAB |
| 2018, Longitudinal safety evaluation of connected vehicles' platooning on expressways", Accident Analysis & Prevention | Rahman., M – University of Central Florida | Platooning | Freeway | DSRC of 300 m (1000 feet) | Safety | All five surrogate measures of safety (standard deviation of speed, time exposed time-to-collision TET, time integrated time-to-collision TTT, time exposed rear-end crash risk index, sideswipe crash risk) improved. Managed-lane CV outperformed all-lane CV platooning. | Simulation | VISSIM |
| 2014, Improving traffic operations using real-time optimal lane selection with connected vehicle technology | Jin, Q. – University of California Riverside | Optimal Lane Selection (OLS) | Freeway | N/A | Safety, Environment | Travel time reduced 3.8%; fuel consumption reduced 2.2%; emissions reduced from 1% to 19% depending on the congestion level. | Simulation | SUMO |
| 2016, How to assess the benefits of connected vehicles? A simulation framework for the design of cooperative traffic management strategies | Gueriau M. – Université de Lyon | Advanced Driver Assistance System (ADAS) | Freeway | N/A | Safety, Mobility | Speed decreased with an homogenization of speeds and headways | Simulation | A multi-agent framework embedded with MovSim |
| 2013, Cooperative Highway Traffic Multiagent Modeling and Robustness Assessment of Local Perturbations | Monteil J. – Université de Lyon | Cooperative Car Following | Freeway | N/A | Mobility | With the cooperative traffic, aggressive lane-changing behavior is reduced, and the stability and homogenization of traffic is achieved. | Simulation | OVRV and MOBIL |

| Year, Title | Author, Leading institute | CV Application | Roadway Type | Comm. Type | Topic Area | Estimated Benefits | Simulation/ Field Test | Modeling Environment |
|---|---|---|---|---|---|---|---|---|
| 2020, Mobility and energy consumption impacts of cooperative adaptive cruise control vehicle strings on freeway corridors | Liu, H.– UC Berkeley | Platooning | Freeway | N/A | Mobility, Environment | Average mpg of 27 was achieved when MPR is 50%; mobility increased 30% with MPR of 10%. | Simulation | MOVES and VT-CPFM models |
| 2021, Energy and flow effects of optimal automated driving in mixed traffic: Vehicle-in-the-loop experimental results | Ard, T. – Clemson University | CV Car following control with MPC | Freeway | V2Sim | Mobility, Environment | Improvement of 30% in energy economy; increased travel time and headway. | Simulation | Automated driving virtual simulation with a physical vehicle (EGO) |
| 2020, Model-free speed management for a heterogeneous platoon of connected ground vehicles | Weng, Y. – U of Michigan | Eco-CACC | Freeway | N/A | Mobility, Environment | Utility improvement over initial speed is 40% in mobility and 2.2% in fuel economy; utility improvement over desired speed is 18.4% in fuel economy but at the cost of 7.5% mobility. | Simulation | military trucks |
| 2015, Efficient vehicle driving on multi-lane roads using model predictive control under a connected vehicle environment | Kamal, M.A.S. – Gunma University | Model Predictive Control (MPC) | Freeway | N/A | Mobility | 6.79% increased velocity and 7.22% fuel economy. | Simulation | MATLAB |
| 2017, Cooperative autonomous driving for traffic congestion avoidance through vehicle-to-vehicle communications | Wang, N. – Fujitsu Laboratories | Altruistic Cooperative Driving (ACD) | Freeway | N/A | Mobility | ACD achieves higher speed efficiency (up to 15%). | Simulation | Traffic simulator implemented in Java |
| 2019, A mixed traffic speed harmonization model with connected autonomous vehicles | Ghiasi, A. – Univerisyt of South Florida | SPD-Harm | Freeway | N/A | Mobility, Environment | Varied benefits from different sensor settings. The savings on throughput, speed STD, fuel consumption, and surrogate safety measures can be up to 1.7%,6.5%,4%, and 17%. | Simulation | VT-Micro |
| 2018, Modeling impacts of cooperative adaptive cruise control on mixed traffic flow in multi-lane freeway facilities | Liu, H. – UC Berkeley | CACC | Freeway | N/A | Mobility | Managed lane (ML) and (vehicle awareness device (VAD) is helpful at low and medium MPR, leading to a capacity improvement ranging from 8% to 23%. | Simulation | NGSIM (VT-CPFM+MOVES) |
| 2021, A Cooperative Platooning Controller for Connected Vehicles | Bichiou, Y. – VTTI | Platooning | Freeway | | Mobility, Environment | Travel time, delay, fuel consumption all dropped for connected automated vehicles (CAVs) and non-connected vehicles, with reduction up to 5%, 9.4%, and 8.17%. | Simulation | INTEGRATION |

A model-free approach was proposed by Weng et al. to incorporate the differences of varied vehicle types and combine the goal of fuel economy and mobility platooning. A utility function was formulated using the Nelder-Mead approach. This approach relies on the communication of instantaneous speed and fuel consumption between vehicles in the platoon with mixed vehicle types. The simulation and experimental results showed that this method was effective in increasing the objective function, with less fuel consumption and higher mobility. However, due to the data-driven nature, convergence required some time [21].

A study conducted by Kamal et al. constructed a model predictive control (MPC) framework to efficiently drive a vehicle on multi-lane roads by enhancing the vehicle's capability in lane change and speed adjustment. MPC uses present state information to predict the future behavior through the explicit use of a process model. The results showed that the MPC generated optimal acceleration of the vehicle and the optimal timing to move to the next lane if long term performance gain was anticipated by predicting surrounding traffic [22].

Wang et al. proposed a strategy called Altruistic Cooperative Driving (ACD), where vehicles that are causing congestion should yield the right of way to other vehicles by slowing down or changing lanes. By defining the vehicle driving conditions into maximum, deadlock, and free-run, the strategy changes the deadlock situation into a free-run or maximum to improve overall efficiency. A simulator was generated using Java for a section of multi-lane road. The results showed that the ACD strategy achieved higher speed efficiency with up to 15% improvement and can perform cooperative driving to resolve deadlock conditions in a timely fashion [23].

A study by Ghiasi et al. designed a CAV-based trajectory-smoothing method to control CAVs upstream of a bottleneck to harmonize traffic, improve fuel-efficiency, and reduce environmental impacts. Four steps were adopted: information update, trajectory prediction, shooting heuristic, and damping control. The results showed that the proposed methodology smoothed CAV movements and harmonized the following human-driven vehicles. Improvements were achieved in throughput, speed variations, fuel consumption and surrogate safety measures [24].

Liu et al. conducted a study to model the impacts of CACC on mixed traffic flow on multi-lane freeway facilities. A modeling framework that adopts a new vehicle dispatching model to generate the high-volume traffic flow was proposed. The framework also ensures realistic CACC vehicle behaviors. By incorporating new lane changing rules and automated speed control algorithms, the proposed modeling framework could further reproduce traffic flow dynamics under the influence of CACC operation strategies. Case studies on four-lane freeway segments with on- and off-ramps were illustrated and the results indicated that their proposed modeling framework could improve system mobility for different MPRs. Specifically, at a 60% or lower MPR, managed lane and vehicle awareness devices were helpful and led to a capacity improvement ranging from 8% to 23%. At a 100% MPR, the freeway capacity was 90% higher than the base case [25].

Bichiou et al. proposed an input minimal platooning controller. This controller incorporates various dynamic and kinematic constraints, such as acceleration, velocity, and collision avoidance constraints. The authors tested the controller using a calibrated real network—freeways in downtown Los Angeles—in a simulation environment utilizing the INTEGRATION microscopic simulation software. The results showed that a significant reduction in system-wide travel time, delays, and fuel consumption was achievable. It was also observed that, although the controller only controlled vehicles in the platoon, all the vehicles in the network benefited from the controller [26].

In summary, CV-enabled applications in uninterrupted traffic flow typically regulate car following and lane changing behavior to minimize fuel consumption, maximize throughput, and improve surrogate crash measures, such as the standard deviation of speed, time to collision, etc. The results varied case by case depending on the different applications, congestion levels, and/or MPRs of CVs. In most cases, the percentage of CVs needed to be above a certain level to achieve significant results. The majority of the previous studies used simulation tools to evaluate the effectiveness of the applications, while field tests were rare due to the scarcity of equipped facilities as well as safety concerns.

### C. Signal-free Intersection Applications

Dresner and Stone were the first to introduce the concept of signal-head-free intersection control [27]. In their approach they considered a first-in first-out intersection control mechanism. Mirheli et al. also designed a signal-head-free intersection control logic using a dynamic programming model. The authors used a stochastic look-ahead technique on a Monte Carlo tree search algorithm to determine the near-optimal actions over time to prevent conflicts. By doing so, the model developed by the authors maximized the intersection throughput. The simulation results showed that travel time was significantly reduced at intersections for different demand patterns. Depending on various demand patterns, the improvement ranged between 59% and 84%, compared to intersections controlled by fixed-time and fully-actuated traffic signals [28].

Similarly, Zohdy et al. developed the intersection CACC system to regulate the flow of traffic proceeding through signal-free intersections [29-31]. This work was then extended to optimize the flow of traffic proceeding through a roundabout [32]. Work by Elhenawy et al. investigated the use of game theory to optimize vehicle movements through the intersection [33]. Bichiou and Rakha developed a signal-head free offline controller in [34] and extended it to real-time vehicle trajectory optimization considering dynamic constraints to enhance the mobility of intersections. The results showed that the proposed algorithm outperformed other intersection control strategies by not only producing lower delay but also decreasing vehicle fuel consumption and $CO_2$ emissions [35]. In summary, CACC and i-CACC are efficient in decreasing delays and fuel consumptions comparing to conventional signal controls, abandoning the concepts of intersection control by optimizing the scheduling and movement of vehicles traversing an intersection. Table III lists the studies for signal-free intersections. The existing studies were mainly conducted in a simulation environment.

### D. Signalized Roadway Applications

Numerous studies have attempted to quantify the benefits of V2X enabled applications along signalized roadways, as summarized in Table IV. Some of the previous studies tested V2X-enabled applications along arterials with one or more signalized intersections, where there are traffic controls that force vehicles to decelerate, stop, and/or accelerate at the onset, during, or at the conclusion of a red indication. By using the data transmitted from other vehicles and infrastructures, the V2X-enabled applications can guide vehicles through intersections with minimum delays, emissions, and without colliding into other vehicles by sharing Signal Phasing and Timing (SPaT) data. Table IV summaries these studies.

Early work on GLOSA or eco-driving in the vicinity of signalized intersections started with the work of Kamalanathsharma and Rakha [36-39]. The authors introduced a dynamic programming approach to deriving the fuel-optimum vehicle trajectory. Yang et al. extended this work by predicting the queue at the intersection approach [40] and considering multiple signalized intersections in deriving the optimum solution. By varying demand level, MPRs, phase splits and offsets, as well as the distances among the consecutive traffic signals, they concluded that a fuel consumption saving of 13.8% can be achieved given a 100% MPR. Combining higher MPRs with shorter phase lengths produced larger fuel savings. Demand levels, traffic signal offset, and traffic signal spacing all affect the results [41]. Later Almannaa et al. tested the controller in a controlled field environment. Three scenarios were tested: normal driving, driving with a speed advisory, and automated Eco-CACC. The field experiment tested four red indication offset values randomly delivered to drivers along an uphill and downhill, totaling 1,563 trips by 32 different participants. The results showed that the proposed Eco-CACC system can reduce fuel consumption significantly; a saving in fuel consumption of 31% and travel time of 9% was achieved. The results also demonstrated that automatic control yields more significant benefits than a human control scenario [42].

A GLOSA system was implemented in a study by Bradaï et al. This system coaches the driver to adapt their vehicle speeds such that the drive can safely proceed through upcoming traffic signals during a green indication. It allows reducing stop times and unnecessary accelerations in urban traffic situations, thereby saving fuel and reducing $CO_2$ emissions. The results obtained show a significant reduction in $CO_2$ emissions. However, the authors also stated that the results were obtained under specially designed circumstances where the trip length was only 1,500 meters and the experiment environment can be considered as a straight line with no other vehicles [43].

An eco-approach application was designed to provide drivers with recommendations to encourage "green" driving while approaching, passing through, and departing intersections in a study conducted by Xia et al. Both simulation experimentation and field operational testing were carried out to demonstrate the eco-approach application and to quantify its potential fuel and $CO_2$ savings. The results showed that a communication platform based on a 4G/LTE C-V2X network link and a cloud-based server infrastructure was effective and sufficient for this kind of application. It was found in both the simulation experiment and the field operational testing that an average of 14% fuel and $CO_2$ savings could be achieved [44].

A study by Wang et al. developed a cluster-wise cooperative eco-approach and departure (Coop-EAD) application for CAVs to reduce energy consumption and compared its performance to existing Ego-EAD applications. Instead of considering CAVs traveling through signalized intersections one at a time, the authors' approach strategically coordinates CAVs' maneuvers to form clusters using various operating modes: initial vehicle clustering, intra-cluster sequence optimization, and cluster formation control. The novel Coop-EAD algorithm is applied to the cluster leader, and CAVs in the cluster follow the cluster leader to conduct EAD maneuvers. A preliminary simulation study for a given scenario showed that, compared to an Ego-EAD (speed and location) application, the proposed Coop-EAD application achieves an 11% reduction in energy consumption, up to an 18% reduction in pollutant emissions, and a 50% increase in traffic throughput, respectively [45].

In a study conducted by Moser et al., a CACC approach using stochastic linear model predictive control strategies was proposed. Both V2V and V2I communication were assumed to be present. The objective of this approach was to minimize the fuel consumption in a vehicle-following scenario. By means of a conditional Gaussian model, the probability distribution of the upcoming velocity of the preceding vehicle was estimated based on current measurements and upcoming traffic signal timings. The evaluation of the controllers showed a significant reduction in vehicle fuel consumption compared to the predecessor while increasing safety and driving comfortably [46].

A study conducted by Bento et al. designed a novel intersection traffic management system for CAVs. The developed intelligent traffic management (ITM) techniques were proved to be successful in reducing the delays and emissions without colliding with each other. The data needed for the ITM system is supported by V2X communication where the vehicle position and speed are exchanged among vehicles and infrastructure. In addition to the savings in $CO_2$ emissions and delays, the authors also found that the improvement is more significant when the traffic is heavier [47].

The study conducted by Wan et al. proposed a speed advisory system (SAS) for pre-timed traffic signals to minimize fuel consumption. They showed that the minimal fuel driving strategy alternates between periods of maximum acceleration, engine shut down, and sometimes constant speed. Instead of using this bang-singular-bang control, they employed a sub-optimal solution without necessarily sacrificing the drivability but with significant improvement in fuel economy. The SAS-equipped vehicles not only improved their own fuel economy, but also benefit other conventional vehicles. A higher MPR generated more fuel savings. The cost in traffic flow and travel time was minimum [48].

A simulation-based case study implemented on a hypothetical four-way single-lane approach intersection under varying congestion conditions showed that the Cooperative Vehicle Intersection Control (CVIC) algorithm significantly improved intersection performance compared to conventional actuated intersection control, with a 99% and 33% reduction in stopped delay and total travel time, respectively. In addition, the CVIC algorithm significantly improved air quality and energy savings: 44% reduction of $CO_2$ and 44% fuel consumption savings [49].

The study by Ho Chaudhuri et al. developed a fuel-efficient control strategy for a group of connected HEVs considering congested traffic and traffic signals that are common in urban road conditions. A higher-level controller develops an optimal velocity profile focusing on minimizing the average tractive energy consumption and reducing red light idling. The lower-level controller tracks the velocity profile obtained from the higher-level controller by optimally splitting power between the vehicle engine and the battery using an adaptive equivalent consumption minimization strategy (ECMS). The simulation results showed that using the proposed method, no vehicle had to stop for red lights. ECMS can track the velocity supplied by the higher level controller perfectly and keeps the state of charge of the battery within 1.5% of the initial battery state of charge [50].

Liu et al. designed a cooperative signal control algorithm utilizing CACC datasets and data collected by traditional fixed traffic sensors. The control strategy proved to be effective with mixed traffic as well as 100% CACC traffic. The average vehicle speed and the average vehicle MPG increased by more than 10% and speed increased by 13% when the CACC MPR was 100%, while the speed increased by 36% and MPG improved by more than 34% when the CACC MPR was 40%. Even with a 0% CACC, the proposed method can generate benefits with speed and MPG improvements of 12.5% and 12.2%, respectively [51].

In a study conducted by Lebre et al., a GLOSA system was developed. The results showed that GLOSA could reduce vehicle emissions, waiting time, and travel times. One feature of this study is that it involved real scenario testing. The authors tested the traffic signal with communication devices that transmitted identity, timestamp, location, and phase information. The data were sent to the V2X equipment, which was composed of an IEEE802.11p complaint WiFi router and an antenna mounted to the traffic light, through an Ethernet connection. In the case when there was only one vehicle, both simulation and the in-field test achieved similar savings in $CO_2$ emissions (11% and 13%). When there was more than one vehicle included (only in simulation) a 5% saving in $CO_2$ emissions and 30% of waiting time saving were observed when the MPR was around 50%. The simulations were conducted with two other networks and the savings in time and $CO_2$ emissions were also significant—for the grid network, a reduction of 10.5% in $CO_2$ emissions and an 80% decrease in waiting time was observed with an MPR of 100%;, while for the Valeo shuttle transit route, the reduction of the travel time was about 2.05% with 50% vehicle MPR and 50% MPR for infrastructure. The saving increased quickly to 16% when the CAV MPR increased to 50% and 60% for infrastructure [52].

Chen and Rakha developed a connected eco-driving controller for BEVs that assists BEVs negotiating signalized intersections by minimizing energy consumption. This controller features multiple realistic constraints, including vehicle acceleration and deceleration behavior, BEV energy consumption behavior, and the intercorrelation of speed, location, signal timing in a CV environment. The results showed that the authors' controller can effectively reduce stop-and-go behavior while at the same time generating a savings of 9.3% in energy consumption and 3.9% in delays [53]. In another study, the same authors developed a HEV Eco-CACC system. The system computes real-time energy-optimized trajectories for HEVs. Within the system, two models are developed: one HEV energy model to compute the instantaneous fuel consumption and one vehicle dynamics model to capture the relationship between speed, acceleration as well as tractive/resistance forces. The results revealed that on an arterial with three signalized intersections, the proposed system produced a 7.4% reduction in energy consumption, a 5.8% reduction in delay, and a 23% reduction in vehicle stops.

Elouni et al. designed an adaptive traffic signal control system using two methods—game-theoretic decentralized and centralized perimeter control—and compared the performance of both systems. The results showed that the Nash Bargaining (NB) traffic signal controller can prevent congestion from building and improve the performance of the entire network. The NB controller outperforms both gating and non-gating controllers with significant reductions in vehicle travel time, emissions, fuel consumption, and delays. Specifically, the decentralized NB controller led to significant reductions of, with or without gating, 21% to 41% in total delay, 40% to 55% in $CO_2$ emission levels, and 12% to 20% fuel consumption [54].

A study by Maile tested a cooperative intersection collision avoidance system using vehicles from five OEMs: Daimler, Ford, GM, Honda and Toyota, among which the GM vehicle contained the full prototype. The intersection at 5th Ave. and El Camino Real in Atherton, CA was used for the testing. Results showed that the system can almost 100% reliably send out warnings and therefore prevent crashes from happening [55].

TABLE III
V2X-ENABLED ARTERIAL APPLICATIONS AT SIGNAL-FREE INTERSECTIONS

| Year, Title | Author, Leading institute | CV Application | Road Type | Comm. type | Topic Area | Estimated Benefits | Simulation/ Field Test | Modeling Environment |
|---|---|---|---|---|---|---|---|---|
| 2005, Multiagent Traffic Management: An Improved Intersection Control Mechanism | Dresner, K. – University of Texas, Austin | | Arterial | N/A | Mobility | Average trip time for the proposed reservation system is always at the optimum level, comparing to stop sign, traffic signal-controlled intersections, as well as the overpass | Simulation | |
| 2018, Development of a signal-head-free intersection control logic in a fully connected and autonomous vehicle environment | Mirheli, A. – State University of New York at Stony Brook | | Arterial | N/A | Mobility | A signal-head-free intersection control logic was developed, and will completely avert incidents and significantly reduce travel time ranging between 59% and 84%. | Simulation | VISSIM |
| 2012, Intersection management for autonomous vehicles using i-CACC | Zohdy, I. – VTTI | CACC, i-CACC | Arterial | N/A | Mobility and Environment | Savings in delay and fuel consumption in the range of 91% and 82% relative to conventional signal control were demonstrated, respectively. | Simulation | INTEGRATION |
| 2012, Game theory algorithm for intersection-based cooperative adaptive cruise control (CACC) systems, | Zohdy, I. – VTTI | CACC | Arterial | N/A | Mobility | The proposed system reduces the total delay relative to a traditional stop control by 35 seconds on average, which corresponds to an approximately 70% reduction in the total delay. | Simulation | |
| 2016, Intersection management via vehicle connectivity: The intersection cooperative adaptive cruise control system concept | Zohdy, I. – VTTI | CACC, i-CACC | Arterial | N/A | Mobility and Environment | Four types of intersection control methods were compared and the results show that the proposed i-CACC system significantly reduces the average intersection delay and fuel consumption level by 90% and 45%, respectively. | Simulation | INTEGRATION |
| 2013, Enhancing Roundabout Operations via Vehicle Connectivity | Zohdy, I. – VTTI | CACC | Arterial | N/A | Mobility and Environment | The proposed system can reduce delay and fuel consumption up to 80% and 40%, respectively. | Simulation | |

| Year, Title | Author, Leading institute | CV Application | Road Type | Comm. type | Topic Area | Estimated Benefits | Simulation/ Field Test | Modeling Environment |
|---|---|---|---|---|---|---|---|---|
| 2015, An intersection game-theory-based traffic control algorithm in a connected vehicle environment, | Elhenawy, M. – VTTI | CACC | Arterial | N/A | Mobility | The proposed algorithm demonstrated reduction in travel time and delay in the range of 49% to 89% compared to an all-way stop sign control. | Simulation | INTEGRATION |
| 2018, Developing an Optimal Intersection Control System for Automated Connected Vehicles | Bichiou, Y. – VTTI | N/A | Arterial | N/A | Mobility and Environment | The model developed in this study optimized the movements of AVs subjected to dynamical constrains and static constrains. The results demonstrate that an 80% reduction in delay is achievable compared with the best of these three intersection control strategies (a roundabout, a stop sign, and a traffic signal-controlled intersection), on average. A 42.5% and 40%, reduction in vehicular fuel consumption and $CO_2$ emissions, respectively, were achieved as well. | Simulation | INTEGRATION |
| 2019, Real-time optimal intersection control system for automated/cooperative vehicles | Bichiou, Y. – VTTI | N/A | Arterial | N/A | Mobility and Environment | The control system proposed is proved to be effective in achieving 55% reduction in delay compared to roundabout, four-way stop sign, or a traffic signal-controlled intersection. It also yielded a 43% reduction in fuel consumption and $CO_2$ emissions. | Simulation | INTEGRATION |

TABLE IV
V2X-ENABLED ARTERIAL APPLICATIONS AT SIGNALIZED INTERSECTIONS

| Year, Title | Author, Leading institute | CV Application | Road Type | Comm. type | Topic Area | Estimated Benefits | Simulation/ Field Test | Modeling Environment |
|---|---|---|---|---|---|---|---|---|
| 2013, Fuel–Optimal Vehicle Throttle Control: Model Logic and Preliminary Testing | Kamalanathshar ma, R. - VTTI | N/A | Arterial | N/A | Environment | The proposed variable throttle model provides significant fuel savings. Additional savings of 37% was observed when the Dijkstra minimum path algorithm was applied and savings of 14% was achieved when using an A-start minimum path finding algorithm. | | |
| 2013, Multi-stage dynamic programming algorithm for eco-speed control at traffic signalized intersections | Kamalanathshar ma, R. – VTTI | Eco-CACC | Arterial | N/A | Environment | The model proposed by the authors can save fuel consumption by up to 30% in proximity of signalized intersections. | Simulation | INTEGRATION |
| 2014, Leveraging Connected Vehicle Technology and Telematics to Enhance Vehicle Fuel Efficiency in the Vicinity of Signalized Intersections | Kamalanathshar ma, R. – VTTI | Eco-CACC | Arterial | N/A | Environment | The proposed trajectory optimization using a moving horizon dynamic programming approach was calibrated and tested on 30 top-sold vehicles and the results showed saving in fuel consumption ranging from 5% to 30%. | Simulation | |
| 2012, Agent-based modeling of Eco-Cooperative Adaptive Cruise Control systems in the vicinity of intersections | Kamalanathshar ma, R. – VTTI | Eco-CACC | Arterial | N/A | Environment | The model proposed by the authors can save fuel consumption by up to 30% in proximity of signalized intersections. | Simulation | INTEGRATION |
| 2016, Eco-Cooperative Adaptive Cruise Control at Signalized Intersections Considering Queue Effects | Yang, H. – VTTI | Eco-CACC | Arterial | N/A | Environment | The results showed that a fuel savings of up to 40% can be achieved with a 10% MPR. Multiple lane approach requires a higher MPR compared to a single lane approach to generate significant results. | Simulation | INTEGRATION |
| 2021, Eco-Driving at Signalized Intersections: A Multiple Signal Optimization Approach | Yang, H. – VTTI | Eco-CACC | Arterial | N/A | Environment | An eco-driving system that computes a fuel-optimized trajectory was proposed. Using SPaT data communicated from downstream, the eco–driving system can produce a reduction of 13.8% in fuel consumption with an MPR of 100%. | Simulation | INTEGRATION |
| 2019, Field implementation and testing of an automated eco-cooperative adaptive cruise control system in the | Almannaa, M. – VTTI | Eco-CACC | Arterial | N/A | Mobility, Environment | Travel time was reduced by 9% (downhill 8.1%, uphill 9.9%) and fuel consumption by 31% (downhill saving 38.4% uphill saving 22.6%). | Simulation and Field Test | INTEGRATION |

| Year, Title | Author, Leading institute | CV Application | Road Type | Comm. type | Topic Area | Estimated Benefits | Simulation/ Field Test | Modeling Environment |
|---|---|---|---|---|---|---|---|---|
| vicinity of signalized intersections. | | | | | | | | |
| 2016, A Green Light Optimal Speed Advisor for Reduced CO 2 Emissions | B. Bradaï – Valeo, France | GLOSA | Arterial | DSRC – V2X | Environment | Cumulative $CO_2$ emissions for a 1500 m distance travel (max speed 50km/h or 70km/h) decreased by 13% | Field test | Detailed algorithm is not included |
| 2012, Field operational testing of eco–approach technology at a fixed–time signalized intersection | Xia, H. – UC Riverside | Eco-approach and departure | Arterial | CV2X - LTE | Environment | An average of 14% fuel and $CO_2$ savings can be achieved both for simulation and the field test. Travel time decreased by 0.96%. | Field test, Simulation | Paramics |
| 2018, Cluster-wise cooperative eco-approach and departure application for connected and automated vehicles along signalized arterials | Wang, Z. – UC Riverside | Coop-EAD | Arterial | N/A | Environment | Throughput improved by 50%, energy consumption decreased by 11%, and emissions were reduced up to 20%. | Simulation | MATLAB/Simulink and MOVES |
| 2015, Cooperative adaptive cruise control applying stochastic linear model predictive control strategies | Moser, D. - Institute for Design and Control of Mechatronical Systems at the Johannes Kepler, University of Linz, Austria | CACC | Arterial | N/A | Environment | A CACC control approach was proposed and the fuel consumption in a vehicle–following scenario was minimized. | Simulation | |
| 2019, A study of the environmental impacts of intelligent automated vehicle control at intersections via V2V and V2I communications | Bento, L. – Instituto Politécnico de Leiria | Intelligent Traffic Management techniques | Arterial | N/A | Safety, Environment | The Intelligent Traffic Management techniques reduced the $CO_2$ emissions significantly, and the benefit increased with the demands | Simulation | INTEGRATION |
| 2016, Optimal speed advisory for connected vehicles in arterial roads and the impact on mixed traffic | Wan, N. – Clemson University | Speed Advisory System | Arterial | N/A | Environment and Mobility | A sub-optimal solution was proposed to avid bang-bang control. Much fewer stops and smoother trajectories were obtained with fewer less consumption | Simulation | Paramics |
| 2012, Development and Evaluation of a Cooperative Vehicle Intersection Control Algorithm Under the Connected Vehicles Environment | Lee, J. – University of Virginia | Cooperative Vehicle Intersection Control (CVIC) | Arterial | N/A | Mobility, Environment | 99% and 33% of stop delay and total travel time reductions (comparing to actuated intersection control); 44% reductions of CO 2 and 44% savings of fuel consumption . | Simulation | VISSIM |

| Year, Title | Author, Leading institute | CV Application | Road Type | Comm. type | Topic Area | Estimated Benefits | Simulation/ Field Test | Modeling Environment |
|---|---|---|---|---|---|---|---|---|
| 2016, Hierarchical control strategies for energy management of connected hybrid electric vehicles in urban roads | HomChaudhuri, B. - Clemson University | | Arterial | N/A | Environment | Decreased delays significantly by eliminating stops for red lights. The fuel efficiency improved about 50% and $CO_2$ emission decreased about 40% | Simulation | HEVs |
| 2019, Traffic signal control by leveraging Cooperative Adaptive Cruise Control (CACC) vehicle platooning capabilities | Liu, H. – UC Berkeley | CACC - Arterial | Arterial | N/A | Mobility, Environment | Fuel efficiency increased more than 10% and speed increased by 13% when MPR was 100% (36% increase in speed and 34% MPG if MPR was 40%).67% of capacity increase was observed for major approach and 49% was observed for the minor approach. | Simulation | NGSIM/PATH |
| 2015, Real scenario and simulations on GLOSA traffic light system for reduced CO2 emissions, waiting time and travel time | Lebre, Marie–Ange – VALEO, France | GLOSA | Arterial | DSRC | Mobility, Environment | 10% reduction in $CO_2$ when MPR was 100%; 5% reduction in $CO_2$ and 30% reduction in waiting time was observed (with 40% vehicle MPR and 50% infrastructure MPR). | Field test, Simulation | SUMP |
| 2020, Battery Electric Vehicle Eco-Cooperative Adaptive Cruise Control in the Vicinity of Signalized Intersections | Chen, H. – VTTI | Eco-CACC | Arterial | N/A | Mobility, Environment | A saving of 9.3% in energy consumption and 3.9% in vehicle delays were observed. | Simulation | INTEGRATION |
| 2021, Adaptive Traffic Signal Control: Game-Theoretic Decentralized vs. Centralized Perimeter Control | Elouni, M. – VTTI | Decentralized Nash Bargaining Traffic Controller (DNB) | Arterial | N/A | Mobility, Environment | The reductions, with or without gating, are (in average): travel time between 21% to 41%, in total delay between 40% to 55%, and in emission levels ($CO_2$) and fuel consumption between 12% to 20%. | Simulation | INTEGRATION |
| 2009, Cooperative Intersection Collision Avoidance System for Violations (CICAS-V) for Prevention of Violation-Based Intersection Crashes | Maile, D. – Mercedes-Benz Research & Development | CICAS-V | Arterial | DSRC | Safety | The system can reliably send out warning messages 100% of the time and therefore prevent crashes from happening. | Field test | |

In summary, V2X-enabled applications in interrupted traffic flow typically involve a speed advisory system where vehicles are advised to go through a traffic signal-controlled intersection to avoid unnecessary deceleration, acceleration or stops. By adjusting the speeds of vehicles upstream of traffic signals, these applications can help vehicles go through the intersection with minimum interruption and fuel consumption. The existing studies were mainly conducted in a simulation environment. Most of them studied conventional ICEVs. Some tested other vehicle types, such as BEVs, or trucks and buses. Existing studies illustrated the promising results of using V2X-enabled applications in terms of improvements in environment, mobility, and safety.

### III. MODELING THE MUTUAL DEPENDENCIES OF THE TRANSPORTATION AND COMMUNICATION SYSTEMS

This section briefly describes the work on the modeling of the interdependencies of the transportation and communication systems as it relates to the evaluation of V2X applications, as summarized in Table V.

Veins, one of the well-known frameworks in this category, integrates SUMO, a traffic simulator, and OMNET++, a communication simulator, using the TraCI interface. TraCI is a messaging standard that uses the Transmission Control Protocol (TCP) connections to share messages across the two simulators. One advantage of using TraCI is that it allows for bidirectional coupling of the two simulators. However, a key shortcoming of Veins is that it is unable to model large-scale networks. Veins by default does not support C-V2X mode 4; however, OpenCV2X [56] has extended Veins to support C-V2X modeling. Specifically, OpenCV2X implemented the C-V2X standard in SimuLTE, which is built on top of OMNET++ to support LTE communication [57].

Another integrated simulator was developed which entailed integrating the SUMO traffic and NS-3 communication simulators. The authors extended NS-3 to support C-V2X release 14. The authors did not mention how the integration between NS-3 and SUMO was done, but we assume that they used the TraCI interface as was done in Veins because TraCI is the external interface for SUMO developed by the SUMO authors [58].

In two studies conducted by Elbery et al., the authors developed an integrated simulator, VNetIntSim, which combined the OPNet communication simulator with the INTEGRATION traffic simulator. The integrated framework was used to demonstrate the impact of mobility parameters (traffic stream speed and density) on the communication performance through different applications including the File Transfer Protocol using TCP and Voice over Internet Protocol based on the User Datagram Protocol [59, 60].

Hoque et al. attempted to address the scalability problem by parallelizing one or both simulators to speed up the execution time and thus support simulation of large-scale road networks with hundreds of thousands of vehicles. The authors developed an Integrated Distributed Connected Vehicle Simulator (IDCVS) by incorporating hardware-in-the-loop simulation with the integration of SUMO and OMNET++. The authors provided a partitioning heuristic algorithm that partitions the complex traffic network into two sets of partitions: one for SUMO and one for OMNET++. The tool was then used to model DSRC for connected vehicles considering different levels of market penetration [61].

The Vehicular Network Simulator (VNS); which integrates the traffic simulator DIVERT 2.0 and the communication system simulator, NS-3; falls into the third category of methods. VNS supports the 802.11p communication standard as it was developed before the release of the C-V2X LTE standard. VNS differs from the previous work we discussed in the way that NS-3 and DIVERT are integrated. Since both simulators are developed in the same programming language (C++), they were put into one executable environment instead of having the two executing programs communicating with each other. Although the two simulators share the same execution environment, they still communicate using TCP connections through the network integration module. In VNS, at each simulation time step, the traffic simulator is run first followed by the communication simulator. The communication simulator has node entities that are mapped to the vehicle entities in the traffic simulator. Each node entity has access to its corresponding vehicle entity. One last difference in VNS is the adaptation of the NS-3 network simulator. The authors adapted the implementation of NS-3 to support large-scale simulations. They applied the concept of nearest neighbors and the locality of vehicle position updates by using QuadTrees to accelerate the performance of the NS-3 communication simulator. However, the integrated simulator is still constrained by the computational speed of NS-3 [62].

Elbery and Rakha developed a network simulator that simultaneously models the traffic and communication systems. Unlike the previous work, the communication system abstraction is not a network simulator, but instead is an analytical model, which allows for large-scale modeling of CV applications. The authors developed an analytical communication model for the DSRC Media Access Control layer protocol that estimates the packet drop probability and delay using a Markov chain and a queuing model. They implemented the analytical model in the INTEGRATION microscopic traffic simulator and tested the integrated framework using a dynamic eco-routing application [63, 64].

TABLE V
COMPARISON OF INTEGRATED TRAFFIC AND COMMUNICATION SIMULATORS

| Year, Title | Author - Institute | Integrated Simulator | Simulation Scale | Network Simulator | Comm. Standard | Vehicle Positions | Simulator Coupling | Spatial Analysis | Modeling Environment |
|---|---|---|---|---|---|---|---|---|---|
| 2021, INTEGRATION Large-Scale Modeling Framework of Direct Cellular Vehicle-to-All (C-V2X) Applications | Farag, M – VTTI | INTEGRATION | Large scale | Analytical model | Direct C–V2X | Grid cell and update index | Dynamic interval | Yes | INTEGRATION |
| 2021, VNS: An Integrated Framework for Vehicular Networks Simulation | Fernandes, R. – University of Porto, Portugal | VNS | Large scale | NS-3 | DSRC (802.11b) | Quad Tree | Fixed interval | No | DIVERT |
| 2019, OpenCV2X Mode 4 A Simulation Extension for Cellular Vehicular Communication Networks | McCarthy, B – University College Cork Ireland | VEINS | Small scale | OMNET++ | IEEE 802.11b | NA | Fixed interval | No | SUMO |
| 2011, Bidirectionally Coupled Network and Road Traffic Simulation for Improved IVC Analysis | Sommer, C. – University of Erlangen | Open Source C-V2X | Small scale | OMNET++ | C-V2X | NA | Fixed interval | No | SUMO |
| 2019, Performance Analysis of C-V2X Mode 4 Communication Introducing an Open–Source C–V2X Simulator | Eckermann, F – TU Dortmund University | Open Source C-V2X | Medium Scale | NS-3 | C-V2X | NA | Fixed interval | No | SUMO |
| 2015, VNetIntSim - An Integrated Simulation Platform to Model Transportation and Communication Networks/2015, An Integrated Architecture for Simulation and Modeling of Small- and Medium-Sized Transportation and Communication Networks | Elbery, A – VTTI | VNetIntSim | Medium scale | OPNET | IEEE 82.11g | NA | Fixed interval | No | INTEGRATION |
| 2019, VNetIntSim – An Integrated Simulation Platform to Model Transportation and Communication Networks/2019, Large-Scale Modeling of VANET and Transportation Systems | Elbery, A – VTTI | INTEGRATION | Large scale | Analytical model | DSRC (IEEE 802.11p) | NA | Fixed interval | No | INTEGRATION |

| Year, Title | Author - Institute | Integrated Simulator | Simulation Scale | Network Simulator | Comm. Standard | Vehicle Positions | Simulator Coupling | Spatial Analysis | Modeling Environment |
|---|---|---|---|---|---|---|---|---|---|
| 2019, Parallel Closed-Loop Connected Vehicle Simulator for Large-Scale Transportation Network Management: Challenges, Issues, and Solution Approaches | Hoque, M.A. – East Tennessee State University | IDCVS | Large scale | OMNET++ | DSRC (IEEE 802.11p) | NA | Fixed interval | Yes | SUMO |

TABLE VI
ROAD NETWORK ARCHITECTURE AND SIMULATION TIME

| Year, Title | Author – Institute | Road Network | Simulation Time | Number of Vehicles | Execution Time |
|---|---|---|---|---|---|
| 2021, INTEGRATION Large-Scale Modeling Framework of Direct Cellular Vehicle-to-All (C-V2X) Applications | Farag, M – VTTI | Downtown LA. Area 133 $km^2$. A total of 1624 nodes, 3556 links, and 457 traffic signals. | 1.8 h | 145,000 vehicles with a maximum of 30,000 concurrent vehicles | 1.5 h |
| 2019, VNetIntSim - An Integrated Simulation Platform to Model Transportation and Communication Networks/2019, Large-Scale Modeling of VANET and Transportation Systems | Elbery, A – VTTI | Downtown LA. Area 133 $km^2$. 1625 nodes, 3561 links, and 459 traffic signals (42 RSUs). | 8.3 h | 563,626 vehicles with a maximum of 30,000 concurrent vehicles | 8.3 h |
| 2021, VNS: An Integrated Framework for Vehicular Networks Simulation | Fernandes, R. – University of Porto, Portugal | Road network of city of Porto. | 40 min | 130,000 vehicles with a maximum of 15,000 concurrent vehicles | 7 h |
| 2015, VNetIntSim - An Integrated Simulation Platform to Model Transportation and Communication Networks/2015, An Integrated Architecture for Simulation and Modeling of Small- and Medium-Sized Transportation and Communication Networks | Elbery, A – VTTI | An intersection and four zones. Each zone serves as a vehicle origin and destination location. Each road link is 2 km long. | Not reported | 3000 vehicles with 180 concurrent vehicles | Not reported |
| 2019, Performance Analysis of C-V2X Mode 4 Communication Introducing an Open-Source C-V2X Simulator | Eckermann, F – TU Dortmund University | A 100 m × 100 m intersection, and an urban Manhattan grid scenario as used by 3GPP (750 m × 1299 m). | 30 s | 250 vehicles | Not reported |
| 2011, Bidirectionally Coupled Network and Road Traffic Simulation for Improved IVC Analysis | Sommer, C. – University of Erlangen | A 2700 m six-lane highway section, lane width of 4 m, vehicular speeds of 140 km/h (70 km/h). The inter-vehicle distance of 2.5 s × maximum speed. | Not reported | 200 (380) vehicles in the simulation at its most dense stage | Not reported |
| 2019, OpenCV2X Mode 4 A Simulation Extension for Cellular Vehicular Communication Networks | McCarthy, B – University College Cork Ireland | Single-lane Manhattan Grid with intersections spaced 1 km apart. Grid sizes 5 × 5 roads and 16 × 16 roads. | Not reported | 30 and 1000 vehicles | Not reported |

Continuing the work of Elbery and Rakha, Farag et al. developed an integrated C-V2X and traffic simulation framework to simultaneously model the traffic and communication systems and their bi-directional coupling in a large-scale road network. The authors again used the INTEGRATION traffic simulator and an analytical model to implement the communication features of C-V2X and coupled them together in one execution environment. They scaled up the running time by leveraging a spatial index to accelerate the computation time of the communication model and tested it on the downtown Los Angeles network, modeling a total of approximately 145,000 vehicles [65]. In Table VI, we summarize the various parameters associated with the different simulator studies. This includes the different road networks, simulation times, number of vehicles simulated, and execution time for the various CV application simulators.

In summary, developing a fully integrated traffic and communication simulator is essential for the modeling CV applications. Most of the work integrates the two systems using dedicated simulators in the respective fields, which models low level details of both systems but lacks the capability of large-scale simulations. Recent work uses analytical modeling techniques to substitute one of the dedicated simulators for the benefits of scaling to large scale simulations while not sacrificing significantly on the accuracy of the modeling tool.

## IV. APPLICATION EVALUATIONS CONSIDERING THE COMMUNICATION SYSTEM CONSTRAINTS

This section describes how the tools developed in the previous section are used to assess the performance of various CV applications while capturing the mutual interdependencies of the transportation and communication systems.

Schiegg et al. studied the environmental awareness problem of vehicles on the road, which they called incomplete vehicle perception due to the limited range of the vehicle's on-board sensors. They proposed extending the vehicle's perception by sharing the vehicle's information, collected by the vehicle's sensors, with other vehicles on the road using C-V2X communication technology. The service of sharing information within vehicles on the road is called collective perception. The authors proposed an analytical model to evaluate the performance of the service using the C-V2X Mode 4 standard and used an analytical model to model the C-V2X standard. The authors found that the collective perception service enhanced the information when using C-V2X to share sensors' information. They also found that although C-V2X was useful, more enhancement was needed regarding the latency requirements of vehicle safety applications [66].

In the study by Segata et al., the authors studied the performance of the C-V2X communication standard in the context of platooning applications. The authors investigated the impact of the scheduling algorithm for Mode 4 on the platoon formation using several controllers. Findings suggested that although C-V2X performed very well in terms of Packet Delivery Ratio, it did not perform well in terms of packet loss bursts. The authors showed that the packet loss bursts can hinder C-V2X usefulness in safety applications, and claimed that the packet loss bursts are due to the scheduling algorithm and the half-duplex nature of the C-V2X channel (vehicles cannot receive and transmit at the same time) [67].

A study by Rajab and Miucic investigated the performance of the C-V2X communication standard on the performance of two safety applications: Emergency Vehicle Alert and High Beam Assist. The authors modeled the C-V2X communication standard using empirical data from the Crash Avoidance Metrics Partners performance assessment project to create several Packet Error Ratio (PER) curves for different conditions (ideal, medium, and severe). They smoothed the PER curves using splines applied to the empirical data [68].

A truck platoon application was tested in [69] using C-V2X communication. The authors found that the best scenario is to use C-V2X mode 3 in areas covered with LTE-infrastructure and C-V2X mode 4 in areas not covered to get the best benefits from the platoon (i.e., smaller inter-truck gaps). The authors, however, noted that one critical configuration of the C-V2X mode 4 is the re-selection counter, which must be tuned carefully to achieve good performance [69].

Marco et al. developed an open-source framework for testing V2X applications. They evaluated two applications: emergency vehicle (V2V) and area speed advisory (V2I/V2N) using DSRC and C-V2X communication technologies. They showed that the two applications were valuable and did help the emergency vehicle to maintain high speed while crossing intersections (in the V2V case) and decreased the number of collisions in the area of the speed advisory applications (in the V2I/V2N case) [70].

Mouawad et al. tested and evaluated a cooperative collision avoidance V2I application at urban intersections. The vehicles shared their Lidar sensor measurements (local occupancy maps) with RSUs. The RSUs fused all the messages into a global occupancy map and sent it back to all vehicles in range. The authors found that the configuration for the best performance was a message size of 1,685 bytes for global occupancy maps, which led to the lowest obstacle mis-detection rate and a packet generation rate of 10 Hz when the number of vehicles was less than 70 and 5 Hz otherwise [71].

A C-V2I-based system for collision avoidance was introduced and evaluated in a study by Malinverno et al. The authors evaluated the effectiveness of their collision detection algorithm using Cooperative Awareness Messages that were sent by vehicles and pedestrians using C-V2I. It was not clear what the configuration of the communication protocol was, but the authors mentioned that they used SimuLTE-veins for simulating the C-V2I. The authors assumed an MPR of 100% and tested the location of the application server in two settings: the cloud and at the RSU. The results of their algorithm was not affected by the location of the application server [72].

Video-assisted overtaking maneuver application using image processing and video streaming over C-V2X was introduced in a study by Magalhaes et al. The C-V2X enhanced the performance of the application by having low packet-loss and high video quality; however, it affected the latency. The authors used a C-V2X onboard unit with a Cooperative Awareness Messages frequency of 10 Hz, HARQ enabled, and transmission power of 20 dBm with Line-of-sight assumption [73].

A study conducted by 5GAA analyzed the capability of 3GPP LTE-V2X PC5 (LTE side-link) and IEEE 802.11p (DSRC or ITS-G5) in reducing the number of fatalities and serious injuries. The model used the number of crashes as a baseline number and the fraction of signal delivery reliability, effectiveness of receiving alert/warning message, as well as some other ratios to estimate the number of crashes that could be avoided. The results showed that LET-V2X would avoid a greater number of crashes when compared to 802.11p due to a combination of the superior performance of LTE-V2X and the market led conditions that better favor the deployment of LTE-V2X [74].

Rebbeck et al. concluded that C-V2X enhances road safety and traffic efficiency. The base and equitable 5.9 GHz user scenario appeared to be the be the most beneficial way to deploy C-V2X considering the cost of upgrading the roadside infrastructure. Additional benefits can be achieved if LTE PC5 communication is integrated in smartphones. The study also concluded that the cost of upgrading the in-vehicle C-ITS system would be significant [75].

Beyrouty et al. evaluated seven bundles of C-ITS services that are mature and are expected to be deployed in the short or medium term, including safety-based V2V services, V2I services that deliver the most benefit on motorways, V2I services mostly appliable in urban areas, services intended to provide information regarding parking, services intended to provide traffic and smart routing information, and V2X vulnerable road user protection services. The authors also assessed policy options considering six key themes: privacy, security, interoperability, compliance assessment, continuity, and enabling conditions. They estimated the savings in avoiding crashes, fuel consumption savings, and emission reductions for different policy options [76].

Using the INTEGRATION simulation tool Elbery and Rakha evaluated the mutual impact of an eco-routing transportation application and the V2I communication system using the 802.11p protocol. They tested the eco-routing application performance using different measures of effectiveness for different CV MPRs and congestion levels. The results of their study showed that reasonable fuel savings were achieved at a low-to-medium MPR. However, at high MPRs and high vehicular traffic congestion levels the eco-routing application performance degraded and caused higher fuel consumption and even a gridlock in the road network. This was because of the sub-optimal routes produced by the eco-routing application caused by the high packet drop rate [77, 78]. Alternatively, ignoring the communication system constraints produced benefits that increased as the CV MPR increased. This study clearly demonstrated the need to model the communication system as part of the evaluation of CV applications and demonstrated that without such an integrated modeling framework, erroneous conclusions could be derived.

In summary, V2X-enabled intelligent transportation applications like collective perception, platooning, emergency vehicle alert, area speed advisory, collision avoidance, and video-assisted overtaking maneuvers have been tested in the literature. Despite the promising benefits of V2X technology, there is still more room for enhancement, specifically from the latency perspective.

## V. CONCLUSIONS AND RECOMMENDATIONS FOR FURTHER RESEARCH

This paper reviewed numerous research efforts that attempted to quantify the impacts of various CV applications on the transportation system's mobility, safety, and the environment. Based on the literature review, the following conclusions can be derived:

1. CV applications are highly efficient in improving the transportation system's mobility and safety and minimizing vehicle emissions and fuel consumption levels. Assuming perfect wireless communication among the different components of the CV environment, the mobility, safety, and environmental impacts typically increase as the MPR of CVs increases.

2. The MPR of CVs is a critical factor that affects the performance of CV applications. A minimum MPR was recommended in many studies to reach a significant level of savings in travel time, delays, and emissions. Depending on the types and features of the different applications, the minimum MPR varies from 10% to 40%.

3. The benefits of CV applications generally increase with the increase of MPR if communication constraints are not explicitly accounted for. The benefits, however, do not follow a linear relationship with the MPR. Some applications will start to show effects once a minimum number of CVs are in the traffic flow, but the benefits flatten out after the MPR reaches a certain level. Some applications do not generate benefits until the MPR reaches a certain level; however, the benefits increase accordingly with the increasing percentage of CVs in the traffic flow.

4. Most of the previous research efforts assumed that the communication among different parts of the system is latency and/or error-free, assuming the C-V2X data are accepted and applied by all users without an error. This assumption should be modified in future research. Due to the delays in signal transmission, bandwidth, and varied level of acceptance and cooperation of travelers in the system, the benefits will be lower in real-world situations. The accurate benefits in the real world need to be estimated accordingly accounting for limitations in the communication system.

5. A unique study of an eco-routing application showed that a minimum MPR of about 10% was needed to achieve benefits, with the benefits peaking at a CV MPR of 30–40%. Higher MPRs loaded the 802.11p communication system and at an MPR of 75%, the application produced gridlock because of the loss and latency in the packet transmission. This finding is unique because it demonstrates the importance of including the communication system constraints in the modeling of CV applications.

6. Most previous research efforts were based on microscopic simulations. Very few studies tested CV technologies in the field given the high cost of such testing. Even in those cases with field testing, the testbed typically was a closed environment involving a small sample of CVs (no more than 10), thus not overloading the communication system.

7. The benefits of CV applications vary based on the application type, study design, vehicle type, test location or network type, roadway condition, and utilized energy/emission model. The differences might also be caused by different modeling algorithms, modeling assumptions or parameter settings in the model.

Based on these findings, we recommend the following research directions:

1. Further research is needed to develop a C-V2X application protocol that can define minimum modeling requirements. This could include variables like vehicle type, test location or network configuration, roadway congestion level, and market penetration levels.

2. Modeling tools that capture the interdependency of the transportation and communication systems are critical to the assessment of CV applications. Further research is needed to develop scalable modeling tools that provide a good abstraction of the communication and transportation systems.

3. Validation of simulation-dominant studies using high fidelity modeling tools or real-world empirical data is needed.

4. Further work is needed to quantify the impact of varied MPRs and V2X applications using these integrated modeling tools to identify the most effective parameters and identify bottlenecks in the communication system.

APPENDIX

*A. Abbreviations*

5G – the 5th generation mobile network
ACC - Adaptive Cruise Control
BEV - Battery Electric Vehicle
CACC – Cooperative Adaptive Cruise Control
C-ITS - Cooperative Intelligent Transport System
Coop-EAD - Cooperative Eco-Approach and Departure
CPI - Crash Potential Index
CV - Connected Vehicle
C-V2X - Cellular Vehicle-to-Everything
CVIC - Cooperative Vehicle Intersection Control
CVLLA - CVs and lower-level automation
DSRC - Dedicated Short-Range Communication
ECMS - Adaptive Equivalent Consumption Minimization Strategy
Eco-CAC - Eco-Cooperative Automated Control
Eco-CACC - Eco-Cooperative Adaptive Cruise Control
EPA - Environmental Protection Agency
GLOSA - Green Light Optimal Speed Advisory
HEV - Hybrid Electric Vehicle
ICEV - Internal Combustion Engine Vehicle
IDCVS - Integrated Distributed Connected Vehicle Simulator

ITM - Intelligent Traffic Management
MPR - Market Penetration Rate
NB – Nash Bargaining
PER - Packet Error Ratio
RSU – Roadside Unit
SAS - Speed Advisory System
SPD-HARM – Speed Harmonization
TCP - Transmission Control Protocol
TET - Time Exposed to Collision
TIT - Time Integrated Time to Collision
TNO - Netherlands Organization for Applied Science
TTC - Time-to-Collision
V2C - Vehicle-to-Cloud Communication
V2D - Vehicle-to-Device Communication
V2G - Vehicle-to-Grid Communication
V2N - Vehicle-to-Network Communication
V2P - Vehicle-to-Pedestrian Communication
V2V - Vehicle-to-Infrastructure Communication
V2X - Vehicle-to-Everything Communication
VMT - Vehicle Miles Traveled
VNS - Vehicular Network Simulator
VSL - Variable Speed Limit

ACKNOWLEDGMENT

The authors would like to thank Georgios Efraimidis, Anne-Lise Thieblemont, and Jim Misner of Qualcomm for their input on the manuscript.

REFERENCES

[1] United States Environmental Protection Agency (EPA), "Inventory of U.S. Greenhouse Gas Emissions and Sinks," 2021. [Online]. Available: https://www.epa.gov/sites/default/files/2021-04/documents/us-ghg-inventory-2021-main-text.pdf?VersionId=wEy8wQuGrWS8Ef_hSLXHy1kYwKs4.ZaU

[2] United States Environmental Protection Agency (EPA). "Sources of Greenhouse Gas Emissions." https://www.epa.gov/ghgemissions/sources-greenhouse-gas-emissions (accessed.

[3] blog.rgbsi.com. "7 Types of Vehicle Connectivity." https://blog.rgbsi.com/7-types-of-vehicle-connectivity (accessed September 20, 2021).

[4] K. Ahn, J. Du, M. Farag, and H. Rakha, "Evaluating an Eco-Cooperative Automated Control System (Eco-CAC)," presented at the The 101th Transportation Research Board Annual Meeting, Washington D.C., 2021.

[5] E. Charoniti *et al.*, "Environmental Benefits of C-V2X ", 2020.

[6] A. Olia, H. Abdelgawad, B. Abdulhai, and S. N. Razavi, "Assessing the Potential Impacts of Connected Vehicles: Mobility, Environmental, and Safety Perspectives," *Journal of Intelligent Transportation Systems,* vol. 20, no. 3, pp. 229-243, 2016/05/03 2016, doi: 10.1080/15472450.2015.1062728.

[7] A. Vahidi and A. Sciarretta, "Energy saving potentials of connected and automated vehicles," *Transportation*

*Research Part C: Emerging Technologies,* vol. 95, pp. 822-843, 2018.

[8]     M. S. Rahman, M. Abdel-Aty, J. Lee, and M. H. Rahman, "Safety benefits of arterials' crash risk under connected and automated vehicles," *Transportation Research Part C: Emerging Technologies,* vol. 100, pp. 354-371, 2019/03/01/ 2019, doi: https://doi.org/10.1016/j.trc.2019.01.029.

[9]     K. Ahn, Y. Bichiou, M. Farag, and H. Rakha, "Multi-objective Eco-Routing Model Development and Evaluation for Battery Electric Vehicles," *Transportation Research Record,* 2021, doi: https://doi.org/10.1177/0361198121103159.

[10]    H. M. Abdelghaffar, M. Elouni, Y. Bichiou, and H. Rakha, "Development of a Connected Vehicle Dynamic Freeway Variable Speed Controller," *IEEE Access,* vol. 8, pp. 99219-99226, 2020.

[11]    J. Jang, J. Ko, J. Park, C. Oh, and S. Kim, "Identification of safety benefits by inter-vehicle crash risk analysis usingconnected vehicle systems data on Korean freeways," *Accident Analysis and Prevention,* vol. 144, no. 105675, 2020, doi: https://doi.org/10.1016/j.aap.2020.105675.

[12]    F. Outay, F. Kamoun, F. Kaisser, D. Alterri, and A. Yasar, "V2V and V2I Communications for Traffic Safety and CO2 Emission Reduction: A Performance Evaluation," *Procedia Computer Science,* vol. 151, pp. 353-360, 2019.

[13]    H. Liu, S. E. Shladover, X.-Y. Lu, and X. Kan, "Freeway vehicle fuel efficiency improvement via cooperative adaptive cruise control," *Journal of Intelligent Transportation Systems,* pp. 1-13, 2020, doi: 10.1080/15472450.2020.1720673.

[14]    H. W. Ye Li, Wei Wang, Shanwen Liu, Yun Xiang "Reducing the risk of rear-end collisions with infrastructure-to-vehicle (I2V) integration of variable speed limit control and adaptive cruise control system,," *Traffic Injury Prevention,* vol. 17, no. 6, pp. 597-603, 2016, doi: 10.1080/15389588.2015.1121384.

[15]    M. S. Rahman and M. Abdel-Aty, "Longitudinal safety evaluation of connected vehicles' platooning on expressways," *Accident Analysis & Prevention,* vol. 117, pp. 381-391, 2018/08/01/ 2018, doi: https://doi.org/10.1016/j.aap.2017.12.012.

[16]    Q. Jin, G. Wu, K. Boriboonsomsin, and M. Barth, "Improving traffic operations with real-time optimal lane selection with connected vehicle technology," in *2014 IEEE Intelligent Vehicles Symposium Proceedings,* 8-11 June 2014 2014, pp. 70-75, doi: 10.1109/IVS.2014.6856515.

[17]    M. Guériau, R. Billot, N.-E. E. Faouzi, J. Monteil, F. Armetta, and S. Hassas, "How to assess the benefits of connected vehicles? A simulation framework for the design of cooperative traffic management strategies," *Transportation Research Part C-emerging Technologies,* vol. 67, pp. 266-279, 2016.

[18]    J. Monteil, R. Billot, J. Sau, F. Armetta, S. Hassas, and N.-E. E. Faouzi, "Cooperative Highway Traffic: Multiagent Modeling and Robustness Assessment of Local Perturbations," *Transportation Research Record,* vol. 2391, no. 1, pp. 1-10, 2013/01/01 2013, doi: 10.3141/2391-01.

[19]    H. Liu, X.-Y. Lu, and S. E. Shladover, "Mobility and Energy Consumption Impacts of Cooperative Adaptive Cruise Control Vehicle Strings on Freeway Corridors," *Transportation Research Record,* vol. 2674, no. 9, pp. 111-123, 2020, doi: 10.1177/0361198120926997.

[20]    T. Ard *et al.,* "Energy and flow effects of optimal automated driving in mixed traffic: Vehicle-in-the-loop experimental results," *Transportation Research Part C: Emerging Technologies,* vol. 130, p. 103168, 2021/09/01/ 2021, doi: https://doi.org/10.1016/j.trc.2021.103168.

[21]    Y. Weng *et al.,* "Model-free speed management for a heterogeneous platoon of connected ground vehicles," *Journal of Intelligent Transportation Systems,* pp. 1-15, 2020, doi: 10.1080/15472450.2020.1797506.

[22]    M. A. S. Kamal, S. Taguchi, and T. Yoshimura, "Efficient vehicle driving on multi-lane roads using model predictive control under a connected vehicle environment," in *2015 IEEE Intelligent Vehicles Symposium (IV),* 28 June-1 July 2015 2015, pp. 736-741, doi: 10.1109/IVS.2015.7225772.

[23]    N. Wang, X. Wang, P. Palacharla, and T. Ikeuchi, "Cooperative autonomous driving for traffic congestion avoidance through vehicle-to-vehicle communications," in *2017 IEEE Vehicular Networking Conference (VNC),* 27-29 Nov. 2017 2017, pp. 327-330, doi: 10.1109/VNC.2017.8275620.

[24]    A. Ghiasi, X. Li, and J. Ma, "A mixed traffic speed harmonization model with connected autonomous vehicles," *Transportation Research Part C: Emerging Technologies,* vol. 104, pp. 210-233, 2019/07/01/ 2019, doi: https://doi.org/10.1016/j.trc.2019.05.005.

[25]    H. Liu, X. Kan, S. E. Shladover, X.-Y. Lu, and R. E. Ferlis, "Modeling impacts of Cooperative Adaptive Cruise Control on mixed traffic flow in multi-lane freeway facilities," *Transportation Research Part C: Emerging Technologies,* vol. 95, pp. 261-279, 2018/10/01/ 2018, doi: https://doi.org/10.1016/j.trc.2018.07.027.

[26]    Y. Bichiou, H. A. Rakha, and H. M. Abdelghaffar, "A Cooperative Platooning Controller for Connected Vehicles," presented at the 7th International Conference on Vehicle Technology and Intelligent Transport System, 2021.

[27]    K. Dresner and P. Stone, "Multiagent Traffic Management: An Improved Intersection Control Mechanism," presented at the The Fourth International Joint Conference on Autonomous Agents and Multiagent Systems, Utrecht, The Netherlands, 2005.

[28]    A. Mirheli, L. Hajibabai, and A. Hajbabaie, "Development of a signal-head-free intersection control logic in a fully connected and autonomous vehicle environment," *Transportation Research Part C: Emerging Technologies,* vol. 92, pp. 412-425,

2018/07/01/    2018,    doi: https://doi.org/10.1016/j.trc.2018.04.026.

[29] I. H. Zohdy, R. K. Kamalanathsharma, and H. Rakha, "Intersection management for autonomous vehicles using iCACC," in *Intelligent Transportation Systems (ITSC), 2012 15th International IEEE Conference on*, 2012: IEEE, pp. 1109-1114.

[30] I. H. Zohdy and H. Rakha, "Game theory algorithm for intersection-based cooperative adaptive cruise control (CACC) systems," in *Intelligent Transportation Systems (ITSC), 2012 15th International IEEE Conference on*, 2012: IEEE, pp. 1097-1102.

[31] I. H. Zohdy and H. A. Rakha, "Intersection management via vehicle connectivity: The intersection cooperative adaptive cruise control system concept," *Journal of Intelligent Transportation Systems,* vol. 20, no. 1, pp. 17-32, 2016.

[32] I. Zohdy and H. Rakha, "Enhancing Roundabout Operations via Vehicle Connectivity," presented at the 92nd Transportation Research Board Annual Meeting, Washington DC, January 14-17, 2013.

[33] M. Elhenawy, A. A. Elbery, A. A. Hassan, and H. A. Rakha, "An intersection game-theory-based traffic control algorithm in a connected vehicle environment," in *Intelligent Transportation Systems (ITSC), 2015 IEEE 18th International Conference on*, 2015: IEEE, pp. 343-347.

[34] Y. Bichiou and H. A. Rakha, "Developing an Optimal Intersection Control System for Automated Connected Vehicles," *IEEE Transactions on Intelligent Transportation Systems,* vol. 20, no. 5, p. 9, 2018. [Online]. Available: https://doi.org/10.1109/TITS.2018.2850335.

[35] Y. Bichiou and H. A. Rakha, "Real-time optimal intersection control system for automated/cooperative vehicles," *International Journal of Transportation Science and Technology,* vol. 8, no. 1, pp. 1-12, 2019/03/01/    2019,    doi: https://doi.org/10.1016/j.ijtst.2018.04.003.

[36] R. K. Kamalanathsharma and H. Rakha, "Agent-based modeling of Eco-Cooperative Adaptive Cruise Control systems in the vicinity of intersections," in *15th International IEEE Intelligent Transportation Systems Conference (ITSC)*, 2012: IEEE, pp. 840-845.

[37] R. K. Kamalanathsharma and H. A. Rakha, "Leveraging Connected Vehicle Technology and Telematics to Enhance Vehicle Fuel Efficiency in the Vicinity of Signalized Intersections," *Journal of Intelligent Transportation Systems: Technology, Planning, and Operations,* no. accepted, 2014.

[38] R. K. Kamalanathsharma and H. A. Rakha, "Multi-stage dynamic programming algorithm for eco-speed control at traffic signalized intersections," in *Intelligent Transportation Systems-(ITSC), 2013 16th International IEEE Conference on*, 2013: IEEE, pp. 2094-2099.

[39] R. K. Kamalanathsharma and H. A. Rakha, "Fuel-Optimal Vehicle Throttle Control: Model Logic and

Preliminary Testing," in *20th ITS World Congress*, 2013.

[40] H. Yang, M. V. Ala, and H. A. Rakha, "Eco-Cooperative Adaptive Cruise Control at Signalized Intersections Considering Queue Effects," in *Transportation Research Board 95th Annual Meeting*, 2016, no. 16-1593.

[41] H. Yang, F. Almutairi, and H. Rakha, "Eco-Driving at Signalized Intersections: A Multiple Signal Optimization Approach," *IEEE Transactions on Intelligent Transportation Systems,* vol. 22, no. 5, pp. 2943-2955, 2021, doi: 10.1109/TITS.2020.2978184.

[42] M. H. Almannaa, H. Chen, H. A. Rakha, A. Loulizi, and I. El-Shawarby, "Field implementation and testing of an automated eco-cooperative adaptive cruise control system in the vicinity of signalized intersections," *Transportation Research Part D: Transport and Environment,* vol. 67, pp. 244-262, 2019/02/01/    2019,    doi: https://doi.org/10.1016/j.trd.2018.11.019.

[43] B. Bradaï, A. Garnault, V. Picron, and P. Gougeon, "A Green Light Optimal Speed Advisor for Reduced CO2 Emissions," in *Energy Consumption and Autonomous Driving*: Springer Link, 2016, pp. 141-151.

[44] H. Xia *et al.*, "Field operational testing of eco-approach technology at a fixed-time signalized intersection," presented at the 15th International IEEE Conference on Intelligent Transportation Systems, 2012.

[45] Z. Wang, S. Wu, P. Hao, and M. Barth, "Cluster-wise cooperative eco-approach and departure application for connected and automated vehicles along signalized arterials," *IEEE Transactions on Intelligent Vehicles,* vol. 3,4, pp. 404-413, 2018.

[46] D. Moser, H. Waschol, L. Kirchsteiger, R. Schmied, and L. D. Re, "Cooperative adaptive cruise control applying stochastic linear model predictive control strategies," presented at the 2015 European Control Conference, 2015.

[47] L. C. Bento, R. Parafita, H. A. Rakha, and U. J. Nunes, "A study of the environmental impacts of intelligent automated vehicle control at intersections via V2V and V2I communications," *Journal of Intelligent Transportation Systems,* vol. 23, no. 1, pp. 41-59, 2019/01/02    2019,    doi: 10.1080/15472450.2018.1501272.

[48] N. Wan, A. Vahidi, and A. Luckow, "Optimal speed advisory for connected vehicles in arterial roads and the impact on mixed traffic," *Transportation Research Part C: Emerging Technologies,* vol. 69, pp. 548-563, 2016/08/01/    2016,    doi: https://doi.org/10.1016/j.trc.2016.01.011.

[49] J. Lee and B. Park, "Development and Evaluation of a Cooperative Vehicle Intersection Control Algorithm Under the Connected Vehicles Environment," *IEEE Transactions on Intelligent Transportation Systems,* vol. 13, no. 1, pp. 81-90, 2012, doi: 10.1109/TITS.2011.2178836.

[50] B. HomChaudhuri, R. Lin, and P. Pisu, "Hierarchical control strategies for energy management of connected hybrid electric vehicles in urban roads," *Transportation Research Part C: Emerging Technologies,* vol. 62, pp. 70-86, 2016/01/01/ 2016, doi: https://doi.org/10.1016/j.trc.2015.11.013.

[51] H. Liu, X.-Y. Lu, and S. E. Shladover, "Traffic signal control by leveraging Cooperative Adaptive Cruise Control (CACC) vehicle platooning capabilities," *Transportation Research Part C: Emerging Technologies,* vol. 104, pp. 390-407, 2019/07/01/ 2019, doi: https://doi.org/10.1016/j.trc.2019.05.027.

[52] M.-A. Lebre, F. L. Mouel, E. Menard, A. Garnault, B. Bradai, and V. Picron, "Real scenario and simulations on GLOSA traffic light system for reduced CO2 emissions, waiting time and travel time," presented at the 22nd ITS World Congress, Bordeaux, France, 2015.

[53] H. Chen and H. A. Rakha, "Battery Electric Vehicle Eco-Cooperative Adaptive Cruise Control in the Vicinity of Signalized Intersections," *Energies,* vol. 13, no. 10, p. 2433, 2020. [Online]. Available: https://www.mdpi.com/1996-1073/13/10/2433.

[54] M. Elouni, H. M. Abdelghaffar, and H. A. Rakha, "Adaptive Traffic Signal Control: Game-Theoretic Decentralized vs. Centralized Perimeter Control," *Sensors,* vol. 21, no. 1, 2021, doi: 10.3390/s21010274.

[55] M. Maile and L. Delgrossi, "Cooperative Intersection Collision Avoidance System for Violations (CICAS-V) for Prevention of Violation-Based Intersection Crashes," *Proceedings: International Technical Conference on the Enhanced Safety of Vehicles,* vol. 2009, pp. -, 2009. [Online]. Available: http://dx.doi.org/.

[56] B. McCarthy and A. O'Driscoll, "OpenCV2X Mode 4 A Simulation Extension for Cellular Vehicular Communication Networks," (in English), *Ieee Int Worksh Comp,* 2019. [Online]. Available: <Go to ISI>://WOS:000556150300011.

[57] C. Sommer, R. German, and F. Dressler, "Bidirectionally Coupled Network and Road Traffic Simulation for Improved IVC Analysis," (in English), *Ieee Transactions on Mobile Computing,* vol. 10, no. 1, pp. 3-15, Jan 2011, doi: Doi 10.1109/Tmc.2010.133.

[58] F. Eckermann, M. Kahlert, and C. Wietfeld, "Performance Analysis of C-V2X Mode 4 Communication Introducing an Open-Source C-V2X Simulator," (in English), *Ieee Vts Veh Technol,* 2019. [Online]. Available: <Go to ISI>://WOS:000610542200472.

[59] A. A. Elbery, H. A. Rakha, M. ElNainay, and M. A. Hoque, "VNetIntSim - An Integrated Simulation Platform to Model Transportation and Communication Networks," in *VEHITS,* 2015.

[60] A. Elbery, H. Rakha, M. Y. ElNainay, and M. A. Hoque, "An Integrated Architecture for Simulation and Modeling of Small- and Medium-Sized Transportation and Communication Networks,"

Cham, 2015: Springer International Publishing, in Smart Cities, Green Technologies, and Intelligent Transport Systems, pp. 282-303.

[61] M. A. Hoque, X. Y. Hong, and S. Ahmed, "Parallel Closed-Loop Connected Vehicle Simulator for Large-Scale Transportation Network Management: Challenges, Issues, and Solution Approaches," (in English), *Ieee Intel Transp Sy,* vol. 11, no. 4, pp. 62-77, Win 2019, doi: 10.1109/Mits.2018.2879163.

[62] R. Fernandes, F. Vieira, and M. Ferreira, "VNS: An Integrated Framework for Vehicular Networks Simulation," (in English), *Ieee Vehic Netw Conf,* pp. 195-202, 2012. [Online]. Available: <Go to ISI>://WOS:000315437900027.

[63] A. Elbery, H. Rakha, and M. ElNainay, "Vehicular Communication and Mobility Sustainability: the Mutual Impacts in Large-scale Smart Cities," p. arXiv:1908.08229, 2012. [Online]. Available: https://ui.adsabs.harvard.edu/abs/2019arXiv190808229E

[64] A. Elbery, H. A. Rakha, and M. ElNainay, "Large-Scale Modeling of VANET and Transportation Systems," in *Traffic and Granular Flow '17,* Washington, DC, USA, 2019, pp. 517-526.

[65] M. M. G. Farag, H. A. Rakha, E. A. Mazied, and J. Rao, "INTEGRATION Large-Scale Modeling Framework of Direct Cellular Vehicle-to-all (C-V2X) Applications," *Sensors,* vol. 21, no. 6, p. 2127, 2021. [Online]. Available: https://www.mdpi.com/1424-8220/21/6/2127.

[66] F. Schiegg, N. Brahmi, and I. Llatser, "Analytical Performance Evaluation of the Collective Perception Service in C-V2X Mode 4 Networks," *2019 IEEE Intelligent Transportation Systems Conference (ITSC),* pp. 181-188, 2019.

[67] M. Segata, P. Arvani, and R. Cigno, "A Critical Assessment of C-V2X Resource Allocation Scheme for Platooning Applications," *2021 16th Annual Conference on Wireless On-demand Network Systems and Services Conference (WONS),* pp. 1-8, 2021.

[68] S. Rajab and R. Miucic, "Assessment of Novel V2X Applications Using a Simulation Platform," 2021. [Online]. Available: https://doi.org/10.4271/2021-01-0115.

[69] V. Vukadinovic *et al.*, "3GPP C-V2X and IEEE 802.11p for Vehicle-to-Vehicle communications in highway platooning scenarios," *Ad Hoc Networks,* vol. 74, pp. 17-29, 2018/05/15/ 2018, doi: https://doi.org/10.1016/j.adhoc.2018.03.004.

[70] M. Malinverno, F. Raviglione, C. Casetti, C.-F. Chiasserini, J. Mangues-Bafalluy, and M. Requena-Esteso, "A Multi-stack Simulation Framework for Vehicular Applications Testing," presented at the Proceedings of the 10th ACM Symposium on Design and Analysis of Intelligent Vehicular Networks and Applications, Alicante, Spain, 2020. [Online]. Available: https://doi.org/10.1145/3416014.3424603.

[71] N. Mouawad, V. Mannoni, B. Denis, and A. P. da Silva, "Impact of LTE-V2X Connectivity on Global

Occupancy Maps in a Cooperative Collision Avoidance (CoCA) System," (in English), *Veh Technol Confe*, 2021, doi: 10.1109/VTC2021-Spring51267.2021.9449034.

[72] M. Malinverno, G. Avino, C. Casetti, C. F. Chiasserini, F. Malandrino, and S. Scarpina, "Performance Analysis of C-V2I-based Automotive Collision Avoidance," (in English), *I S World Wirel Mobi*, 2018. [Online]. Available: <Go to ISI>://WOS:000447268400036.

[73] P. Magalhaes, P. M. D'Orey, M. T. Andrade, and F. Castro, "Video-assisted Overtaking System Enabled by C-V2X Mode 4 Communications," (in English), *Ieee Conf Wirel Mob*, 2020. [Online]. Available: <Go to ISI>://WOS:000662108500064.

[74] A. A. 5GAA, "An Assessment of LTE-V2X(PC5) and 802.11p direct communications technologies for improved road safety in teh EU," 2017.

[75] J. S. Tom Rebbeck, Hugues-Antoine Lacour, Andrew Killeen, David McClure, Alain Dunoyer, "Sociao-Economic Benefits of Cellular V2X," Analysys Mason, SBD Automotive, 2017.

[76] E. L. Kareen Beyrouty, Tom Nokes, Charlotte Brannign, Samuel Levin, Marius Biedka, Hannah Figg, Nick Asselin-Miller, Francesca Fermi, Davide Fiorello, Zani Loredana, Ian Skinner, "Support study for imparc assessment of cooperative intelligent transport systems," 2018.

[77] A. Elbery and H. Rakha, "City-Wide Eco-Routing Navigation Considering Vehicular Communication Impacts," *Sensors*, vol. 19, no. 2, p. 290, 2019. [Online]. Available: https://www.mdpi.com/1424-8220/19/2/290.

[78] A. Elbery and H. A. Rakha, "VANET communcation and mobility sustainability," in *Connected and Autonomous Vehicles in Smart Cities*, 2020, p. 30.

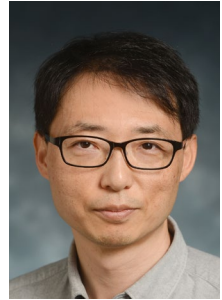

**Kyoungho Ahn** is a Research Scientist in the Center for Sustainable Mobility at the Virginia Tech Transportation Institute. Dr. Ahn received the Ph.D. degree in Civil and Environmental Engineering at the Virginia Tech in 2002. Dr. Ahn has developed and conducted extensive researches as a principal investigator, a co-principal investigator and a faculty investigator in connected and automated vehicle (CAV) applications, vehicle energy and environmental modeling, traffic simulation, traffic flow theory and modeling, environmental impacts on transportation systems, Intelligent Transportation Systems (ITS), and advanced traffic signal control systems. Dr. Ahn has authored/co-authored over 70 refereed publications. The publications were cited more than 4,000 times.

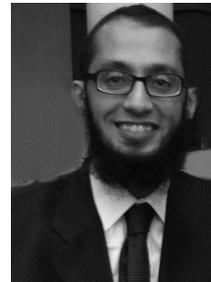

**Mohamed Farag** received the B.Sc. degree (Hons.) in computer engineering from Alexandria University, Alexandria, Egypt, in 2006, the M.Sc. degree in computer science from Arab Academy for science, Technology, and Maritime Transport, Alexandria, Egypt, in 2010, and the Ph.D. degree in computer science from Virginia Tech University, Blacksburg, VA, USA, in 2016. He is currently an Assistant Professor of computer science with the Department of Computer Science, Arab Academy for Science, Technology, and Maritime Transport, and a Research Associate with the Center of Sustainable Mobility, Virginia Tech Transportation Institute. His research interests include intelligent transportation systems, information retrieval, machine learning, large-scale data analysis, and big data.

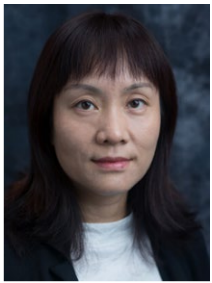

**Jianhe Du** is a Senior Research Associate at Virginia Tech Transportation Institute (VTTI, Blacksburg, VA). She got her Ph.D. from the University of Connecticut in 2005. She is the author/co-author of over 50 papers and reports. She is the principal investigator (PI) and faculty investigator on multiple projects in the area of driving behavior, connected vehicle research, traffic modeling and simulation, Intelligent Transportation Systems (ITS), advanced traffic signal control systems, Geographic Information Systems (GIS) in transportation, and transportation safety modeling. She has extensive experiences on network monitoring and controlling, optimization of traffic signals, traffic simulation, and spatial analysis. She is the registered professional engineer (PE) in the state Virginia. She is also a member of the ITS America and TRB Committee on Geographic Information Science (AED40).

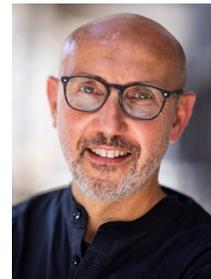

**Hesham A. Rakha** (M'04, SM'18, F'20) received the Ph.D degree from Queen's University, Kingston, Ontario, in 1993. He is currently the Samuel Reynolds Pritchard Professor of Engineering in the Department of Civil and Environmental Engineering and the Department of Electrical and Computer Engineering (Courtesy) at Virginia Tech, and the Director of the Center for Sustainable Mobility at the Virginia Tech Transportation Institute. His research focuses on large-scale transportation system optimization, modeling, and assessment. He works on optimizing transportation system operations, including vehicle routing, developing various network and traffic signal control algorithms, developing freeway control strategies (speed harmonization and ramp metering), and optimizing vehicle motion (lateral and longitudinal control of connected automated vehicles (CAVs)) to enhance their efficiency and reduce their energy consumption while ensuring their safety. Dr. Rakha is a Fellow of the IEEE. He is the recipient of IEEE ITS Outstanding Research Award (2021) from the IEEE Intelligent Transportation Systems Society. He was the author/co-author of six conference best paper awards, namely: 19th ITS World Congress (2012), 20th ITS World Congress (2013), VEHITS (2016), VEHITS (2018), and TRB (2020); received the most cited paper award from the *International Journal of Transportation Science and Technology (IJTST)* in 2018; and received 1st place in the IEEE ITSC 2020 UAS4T Competition. In addition, Dr. Rakha received Virginia Tech's Dean's Award for Outstanding New Professor (2002), the College of Engineering Faculty Fellow Award (2004-2006), and the Dean's Award for Excellence in Research (2007). He is an Editor for *Sensors* (the Intelligent Sensors Section), an Academic Editor for the *Journal of Advanced Transportation*, a Senior Editor for the *IEEE Transactions of Intelligent Transportation Systems*, and an Associate Editor for the *Journal of Intelligent Transportation Systems: Technology, Planning and Operations* and the *SAE International Journal of Sustainable Transportation, Energy, Environment, & Policy*. Furthermore, he is on the Editorial Board of the *Transportation Letters: The International Journal of Transportation Research* and the *International Journal of Transportation Science and Technology*.